\documentclass[a4paper, fleqn, usenatbib, useAMS]{tex/mnras}

\usepackage{amsmath, amssymb}
\usepackage{times, txfonts}

\usepackage[T1]{fontenc}
\usepackage{ae, aecompl}

\usepackage{comment}

\usepackage{graphicx}
\graphicspath{{figs/}}
\DeclareGraphicsExtensions{.pdf, .eps}
\usepackage[multidot]{grffile}

\usepackage{xspace}
\usepackage[usenames, dvipsnames]{xcolor}

\usepackage{alltt}
\usepackage{ulem} 

\newcommand{\etal}{et\thinspace al.\thinspace}
\newcommand{\hii}{H\thinspace\textsc{ii}\xspace}
\newcommand{\bond}{\textsc{bond}\xspace}
\newcommand{\izi}{\textsc{izi}\xspace}
\newcommand{\cloudy}{\textsc{Cloudy}\xspace}

\newcommand{\Ha}{\ifmmode \mathrm{H}\alpha \else H$\alpha$\fi\xspace}
\newcommand{\Hb}{\ifmmode \mathrm{H}\beta \else H$\beta$\fi\xspace}
\newcommand{\neiii}{\ifmmode [\mathrm{Ne}\,\textsc{iii}] \else [Ne~{\scshape iii}]\fi\xspace}
\newcommand{\Neiii}{\ifmmode [\mathrm{Ne}\,\textsc{iii}]\lambda 3869 \else [Ne~{\scshape iii}]$\lambda 3869$\fi\xspace}
\newcommand{\oii}{\ifmmode [\mathrm{O}\,\textsc{ii}] \else [O~{\scshape ii}]\fi\xspace}
\newcommand{\Oii}{\ifmmode [\mathrm{O}\,\textsc{ii}]\lambda 3726 + \lambda 3729 \else [O~{\scshape ii}]$\lambda 3726 + \lambda 3729$\fi\xspace}
\newcommand{\Oiiit}{\ifmmode [\mathrm{O}\,\textsc{iii}]\lambda 4363 \else [O~{\scshape iii}]$\lambda 4363$\fi\xspace}
\newcommand{\heii}{\ifmmode \mathrm{He}\,\textsc{ii} \else He~{\scshape ii}\fi\xspace}
\newcommand{\Heii}{\ifmmode \mathrm{He}\,\textsc{ii}\lambda 4686 \else He~{\scshape ii}$\lambda 4686$\fi\xspace}
\newcommand{\ariv}{\ifmmode [\mathrm{Ar}\,\textsc{iv}] \else [Ar~{\scshape iv}]\fi\xspace}
\newcommand{\Ariv}{\ifmmode [\mathrm{Ar}\,\textsc{iv}]\lambda 4711 + \lambda 4740 \else [Ar~{\scshape iv}]$\lambda 4711 + \lambda 4740$\fi\xspace}
\newcommand{\Niit}{\ifmmode [\mathrm{N}\,\textsc{ii}]\lambda 5755 \else [N~{\scshape ii}]$\lambda 5755$\fi\xspace}
\newcommand{\hei}{\ifmmode \mathrm{He}\,\textsc{i} \else He~{\scshape i}\fi\xspace}
\newcommand{\Hei}{\ifmmode \mathrm{He}\,\textsc{i}\lambda 5876 \else He~{\scshape i}$\lambda 5876$\fi\xspace}
\newcommand{\nii}{\ifmmode [\mathrm{N}\,\textsc{ii}] \else [N~{\scshape ii}]\fi\xspace}
\newcommand{\niis}{\ifmmode [\mathrm{N}\,\textsc{ii}]_\mathrm{S} \else [N~{\scshape ii}]$_\mathrm{S}$\fi\xspace}
\newcommand{\Nii}{\ifmmode [\mathrm{N}\,\textsc{ii}]\lambda 6584 \else [N~{\scshape ii}]$\lambda 6584$\fi\xspace}
\newcommand{\oiii}{\ifmmode [\mathrm{O}\,\textsc{iii}] \else [O~{\scshape iii}]\fi\xspace}
\newcommand{\oiiis}{\ifmmode [\mathrm{O}\,\textsc{iii}]_\mathrm{S} \else [O~{\scshape iii}]$_\mathrm{S}$\fi\xspace}
\newcommand{\Oiii}{\ifmmode [\mathrm{O}\,\textsc{iii}]\lambda 5007 \else [O~{\scshape iii}]$\lambda 5007$\fi\xspace}
\newcommand{\sii}{\ifmmode [\mathrm{S}\,\textsc{ii}] \else [S~{\scshape ii}]\fi\xspace}
\newcommand{\Sii}{\ifmmode [\mathrm{S}\,\textsc{ii}]\lambda 6716 + \lambda 6731 \else [S~{\scshape ii}]$\lambda 6716 + \lambda 6731$\fi\xspace}
\newcommand{\ariii}{\ifmmode [\mathrm{Ar}\,\textsc{iii}] \else [Ar~{\scshape iii}]\fi\xspace}
\newcommand{\Ariii}{\ifmmode [\mathrm{Ar}\,\textsc{iii}]\lambda 7135 \else [Ar~{\scshape iii}]$\lambda 7135$\fi\xspace}
\newcommand{\siii}{\ifmmode [\mathrm{S}\,\textsc{iii}] \else [S~{\scshape iii}]\fi\xspace}
\newcommand{\Siii}{\ifmmode [\mathrm{S}\,\textsc{iii}]\lambda 9069 \else [S~{\scshape iii}]$\lambda 9069$\fi\xspace}

\newcommand{\rOii}{ \ifmmode [\mathrm{O}\,\textsc{ii} ]\lambda 3726/3729 \else [O~{\scshape  ii}]$\lambda 3726/3729$\fi\xspace}
\newcommand{\rOiii}{\ifmmode [\mathrm{O}\,\textsc{iii}]\lambda 4363/5007 \else [O~{\scshape iii}]$\lambda 4363/5007$\fi\xspace}
\newcommand{\rAriv}{\ifmmode [\mathrm{Ar}\,\textsc{iv}]\lambda 4740/4711 \else [Ar~{\scshape iv}]$\lambda 4740/4711$\fi\xspace}
\newcommand{\rNii}{ \ifmmode [\mathrm{N}\,\textsc{ii} ]\lambda 5755/6584 \else [N~{\scshape  ii}]$\lambda 5755/6584$\fi\xspace}
\newcommand{\rSiii}{\ifmmode [\mathrm{S}\,\textsc{iii}]\lambda 6312/9532 \else [S~{\scshape iii}]$\lambda 6312/9532$\fi\xspace}
\newcommand{\rSii}{ \ifmmode [\mathrm{S}\,\textsc{ii} ]\lambda 6731/6717 \else [S~{\scshape  ii}]$\lambda 6731/6717$\fi\xspace}

\newcommand{\qh}{$Q({\mathrm{H^{0}}})$\xspace}
\newcommand{\qhe}{$Q({\mathrm{He^{0}}})$\xspace}

\newcommand{\Np}{N$^{+}$}

\newcommand{\Op}{O$^{+}$}
\newcommand{\Opp}{O$^{++}$}

\newcommand{\Nepp}{Ne$^{++}$}

\newcommand{\Arpp}{Ar$^{++}$}

\title[BOND: Bayesian O \& N abundance Determinations] 
      {BOND: Bayesian Oxygen and Nitrogen abundance Determinations 
       in giant H {\Large \textbf{II}} regions 
       using strong and semi-strong lines}

\author[Vale Asari \etal]
       { N.\ Vale Asari$^{1}$\thanks{e-mail:natalia@astro.ufsc.br},
         G. Stasi\'nska$^{2}$,
         C. Morisset$^{3}$,
         R. Cid Fernandes$^{1}$
         \\
         $^{1}$Departamento de F\'{\i}sica--CFM, Universidade Federal de Santa Catarina, C.P.\ 476, 88040-900, Florian\'opolis, SC, Brazil \\
         $^{2}$LUTH, Observatoire de Paris, PSL Research University, CNRS, Universit\'e Paris Diderot, Sorbonne Paris Cit\'e, \\\hspace{.8em}5 place Jules Janssen, 92195 Meudon, France \\
         $^{3}$Instituto de Astronom\'{\i}a, Universidad Nacional Aut\'onoma de M\'exico \\
       }

\date{Accepted \dots. Received \today; in original form \dots}

\pubyear{2015}

\begin{document}

\label{firstpage}
\pagerange{\pageref{firstpage}--\pageref{lastpage}}

\maketitle

\begin{abstract}

  We present \bond, a Bayesian code to simultaneously derive oxygen and
  nitrogen abundances in giant \hii regions. It compares observed
  emission lines to a grid of photoionization models without assuming
  any relation between O/H and N/O. Our grid spans a wide range in O/H,
  N/O and ionization parameter $U$, and covers different starburst ages
  and nebular geometries.  Varying starburst ages accounts for
  variations in the ionizing radiation field hardness, which arise due
  to the ageing of \hii regions or the stochastic sampling of the
  initial mass function.  All previous approaches assume a strict
  relation between the ionizing field and metallicity. The other novelty
  is extracting information on the nebular physics from
  \textit{semi-strong} emission lines. While strong lines ratios alone
  (\oiii/\Hb, \oii/\Hb and \nii/\Hb) lead to multiple O/H solutions, the
  simultaneous use of \ariii/\neiii allows one to decide whether an \hii
  region is of high or low metallicity.  Adding \hei/\Hb pins down the
  hardness of the radiation field. We apply our method to \hii regions
  and blue compact dwarf galaxies, and find that the resulting N/O vs
  O/H relation is as scattered as the one obtained from the
  temperature-based method. As in previous strong-line methods
  calibrated on photoionization models, the \bond O/H values are
  generally higher than temperature-based ones, which might indicate the
  presence of temperature fluctuations or kappa distributions in real
  nebulae, or a too soft ionizing radiation field in the models.

\end{abstract}


 
\section{Introduction}
\label{intro}

Thanks to their conspicuous emission lines, giant \hii regions are used
as indicators of the chemical composition of the interstellar medium in
galaxies, and have permitted important advances in our understanding of
the chemical evolution of galaxies (see e.g.\
\citealp{Esteban.etal.2004a} and references therein). While the
so-called temperature-based abundance determinations, which require the
measurement of weak auroral lines to measure the electron temperatures,
are commonly considered the most reliable, strong-line methods have
become increasingly popular since the pioneering studies by
\citet{Pagel.etal.1979a} and \citet{Alloin.etal.1979a} because they can
also be applied for distant galaxies.

Strong-line methods involve some restrictions, though: They assume that
giant \hii\ regions form a one (or two) parameter(s) family and they
need to be calibrated. Calibration can be done via a subsample of
objects with temperature-based abundances or using a grid of
photoionization models. The first method is potentially biased, since
calibration samples are likely to have different properties than the
samples one wishes to study. In particular, they are biased against
objects having intrinsically weak auroral lines. Calibrations based on
grids of photoionization models do not have this problem (assuming the
models cover all the combinations of important parameters that are
encountered in nature and are realistic enough).

A large number of calibrations have been proposed. If we label the
methods by their abundance indicators adopting the notation O3N2 for
\oiii/\nii \footnote{In the entire paper the notations \nii, \oii,
  \oiii, \neiii, \sii, \siii, \ariii, \ariv, and \hei\ stand for \Nii,
  \Oii, \Oiii, \Neiii, \Sii, \Siii, \Ariii, \Ariv, and \Hei\
  respectively, \oiiis for $[\mathrm{O}\,\textsc{iii}]\lambda$4959 +
  $\lambda$5007, and \niis for $[\mathrm{N}\,\textsc{ii}]\lambda$6548 +
  $\lambda$6584 (where the subscript S is used to denote a sum of
  emission lines).}, O23 for ($\oii+\oiiis)/\Hb$, N2Ha for \nii/\Ha,
etc., the most popular ones are: O23 \citep{Pagel.etal.1979a}, O3N2
\citep{Alloin.etal.1979a}, O23--O3O2 \citep{McGaugh.1991a}, N2Ha
\citep{StorchiBergmann.Calzetti.Kinney.1994a}, S23
\citep{Vilchez.Esteban.1996a}. All these methods with their numerous
calibrations (here we have quoted the pioneering ones) give very
different outcomes (see e.g. \citealp{Kewley.Ellison.2008a} for a
comparison of the results).

Apart from this problem of leading to discrepant results, strong-line
methods face two important issues. One is that factors other than just
the metallicity and the ionization parameter influence the strength of
the strong lines emitted by giant \hii regions. This potentially leads
to incorrect inferences when applying the methods to compare different
samples \citep[see][]{Stasinska.2010a}. The second issue is that there
are two regimes where the intensities of the strong oxygen lines used
for abundance determinations have the same value with respect to \Hb:
the low-metallicity and the high-metallicity regimes. To resolve this
bimodality, one uses the intensity of the strong nitrogen line since, in
the astrophysical context, the N/O ratio is a function of
metallicity. This procedure, a priori reasonable, is however not totally
secure since the frontier between the two regimes is fuzzy. For example,
\citet{McGaugh.1994a} adopts $\log\, \nii/\oii > -1$ while
\citet{Kewley.Ellison.2008a} adopt $> -1.2$ for the high-metallicity
regime. The difference between these two values may appear insignificant
but can lead to somewhat different conclusions on metallicity trends
within and among galaxies. In addition, this procedure does not allow
one to pin down objects with pathological N/O ratios -- which may be
particularly interesting for unveiling peculiarities in the star-forming
histories of galaxies \citep{Molla.Gavilan.2010a}.  Apart from peculiar
N/O ratios, the N/O vs O/H relation may be different at high redshifts,
which would systematically bias metallicity measurements based on the
local N/O vs O/H relation for high-redshift galaxies.  Of course,
methods using directly N/O as an indicator of the oxygen abundance
(e.g.\ the \nii/\Ha method proposed by
\citealp{StorchiBergmann.Calzetti.Kinney.1994a} or the \nii/\oii method
proposed by \citealp{Kewley.Dopita.2002a}) present the same drawback.

In this paper, we show that using the semi-strong lines \neiii, \ariii
and \Hei, in conjunction with the classical strong lines, it is possible
to estimate with reasonable accuracy both the oxygen and the nitrogen
abundance in giant \hii regions without any prior assumption on the N/O
ratio and without the implicit priors of classical strong line methods
regarding the ionizing radiation field. The intensities of these
semi-strong lines have been listed in many papers reporting on deep
spectroscopy of giant \hii regions, so those lines must be present in
the spectra for which only the most common strong lines have had their
intensities published.

We construct a finely meshed grid of photoionization models varying not
only O/H and the ionization parameter as has been done before
\citep{McGaugh.1991a, Kewley.Dopita.2002a, Blanc.etal.2015a}, but also
N/O \citep[like][]{PerezMontero.2014a}. We use this grid to estimate the
abundances of O and N in giant \hii\ regions by means of standard
Bayesian inference methods \citep[like][]{Blanc.etal.2015a}. Unlike
\citet{Blanc.etal.2015a}, we do not assume an N/O vs O/H relation, and
explicitly explore variations in N/O.  The main novelties of our
approach are that we consider variations in the hardness of the ionizing
field, and that we extract information from {\it semi-strong} emission
lines in addition to the commonly used strong lines.

The paper is organised as follows. In Section \ref{obs} we present the
spectroscopic data we have collected from the literature to develop and
test our method. In Section \ref{noohobserved} we show two extreme
versions of the N/O versus O/H diagram obtained from these data using a
temperature-based method and using the strong-line method of
\citet*{Pilyugin.Vilchez.Thuan.2010a}. In Section \ref{modelgrid} we
present our grid of photoionization models, built using the code \cloudy
\citep{Ferland.etal.2013a}. In Section \ref{method} we present our
method, and in Section \ref{results} we show our results for the N/O
versus O/H diagram. In Section \ref{summary} we provide a summary and
elaborate on future directions of work. Three appendices complement this
paper. The first one presents a realistic sample of fake sources
constructed by selecting model nebulae from our grid.  The second one
describes a few tests using these fake sources. The third one compares
the abundances derived by \bond\ with those obtained by other published
methods on the same set of observational data.


\section{The observational database}
\label{obs}

\begin{table*}
  \caption{A sample of the data table available for download at
    \url{http://bond.ufsc.br}. Line fluxes, uncertainties and upper
    limits (F, eF and limF) are in units of \Hb,
    and references are labelled in column r as in
    Sections~\ref{sec:spirals} and \ref{sec:sBCDs}.}
  \label{tab:sources}
\begin{tabular}{cccccccccccc}
\hline
id & r & name & F3727 & eF3727 & F3869 & eF3869 & \dots & F7135 & eF7135 & limF4363 & limF5755 \\
\hline
001 & a & NGC 1232 02 & 3.9100 & 0.3300 & -- & -- & \dots & 0.0610 & 0.0090 & 0.0140 & 0.0140 \\
002 & a & NGC 1232 03 & 3.3300 & 0.2500 & 0.2530 & 0.0350 & \dots & 0.0720 & 0.0130 & 0.0260 & 0.0260 \\
003 & a & NGC 1232 04 & 2.0700 & 0.1700 & 0.2910 & 0.0240 & \dots & 0.0820 & 0.0090 & 0.0038 & 0.0038 \\
004 & a & NGC 1232 05 & 2.5300 & 0.1600 & 0.0480 & 0.0060 & \dots & 0.0650 & 0.0050 & 0.0022 & -- \\
\dots & \dots & \dots & \dots & \dots & \dots & \dots & \dots & \dots & \dots & \dots & \dots \\
705 & z & HSS1809+6612 & 2.5600 & 0.1013 & 0.3150 & 0.0177 & \dots & -- & -- & -- & -- \\
706 & z & Mrk259 & 2.3900 & 0.0615 & 0.3230 & 0.0099 & \dots & 0.0820 & 0.0033 & -- & -- \\
707 & z & SBS1428 & 1.8800 & 0.0459 & 0.1870 & 0.0059 & \dots & 0.0600 & 0.0022 & -- & -- \\
708 & z & S1657+575 & 2.1300 & 0.0834 & 0.2430 & 0.0157 & \dots & 0.0740 & 0.0083 & -- & -- \\
\hline
\end{tabular}
\end{table*}

\subsection[Giant H II regions in spiral galaxies]{Giant H {\sevensize II} regions in spiral galaxies}
\label{sec:spirals}

Data on giant \hii regions in spiral galaxies were gathered from recent
medium-resolution high-quality observational studies, mostly with very
large telescopes (Keck, VLT) whose high S/N allowed the measurement of
auroral lines in at least part of the observed samples.  Apart from the
large database from \citet{vanZee.etal.1998a}, all the other works involve
Bresolin as first or second author, which guarantees a certain
homogeneity in the treatment of the data. The following sources were
used (the letters correspond the reference labels in Table~\ref{tab:sources}):

\begin{description}
  \item (a) \citet{Bresolin.etal.2005a}; 
  \item (b) \citet{Bresolin.Garnett.Kennicutt.2004a}; 
  \item (c) \citet{Kennicutt.Bresolin.Garnett.2003a}; 
  \item (d) \citet{vanZee.etal.1998a}; 
  \item (g) \citet{Bresolin.etal.2009b}; 
  \item (i) \citet{Bresolin.2007a}; 
  \item (j) \citet{Bresolin.etal.2010a}; 
  \item (k) \citet{Li.Bresolin.Kennicutt.2013a}; 
  \item (l) \citet{Zurita.Bresolin.2012a}; 
  \item (m) \citet{Bresolin.Kennicutt.RyanWeber.2012a}; 
  \item (n) \citet{Goddard.etal.2011a}; 
  \item (p) \citet{Bresolin.etal.2009a}. 
\end{description}

All these sources give the line fluxes corrected for extinction and the
associated uncertainties. When the intensity of \Ha was not given, it
was assumed to be equal to 2.86 times that of \Hb. Many sources list
only the intensities of lines that are used in the classical strong
lines methods. i.e.\ \oiii, \oii and \nii. For papers giving tables with
the fluxes of all the lines seen in the spectra, we roughly estimated
the upper limits for the intensities of the lines that were not
detected, e.g.\ \ariii or \Oiiit. Upper limits are estimated by taking
twice the lowest uncertainty in measured lines. This, of course, is a
stopgap solution in absence of any direct information from the
observers.
      
      
\subsection{Blue compact galaxies}
\label{sec:sBCDs}     
      
To increase the number of objects at low metallicities, we use the
sample of blue compact galaxies with high quality spectra which were
used by \citet[reference labelled z in
Table~\ref{tab:sources}]{Izotov.Thuan.Stasinska.2007a} to derive the
pregalactic helium abundance. The abundances of O and N have been
recomputed in exactly the same way as for the giant \hii\ regions in
spiral galaxies.


\subsection{Subsamples}
\label{sec:subsamples} 
       
For the needs of this study, we merge the two samples described above,
and then constitute several subsamples.

\begin{enumerate}

\item Sample A is constructed from the entire merged sample by selecting
  all the objects with \oii, \oiii and \nii available. It contains 708
  objects.  The line intensities and associated uncertainties are
  reported in Table~\ref{tab:sources}, available for download from
  \url{http://bond.ufsc.br}.

\item Sample T is the subsample of sample A with available temperature
  measurements from \rOiii\ and/or \rNii; it contains 261 objects.

\item Sample B is the subsample fulfilling the minimum requirements for
  the use of the \bond method, i.e. with available fluxes for
    \oii, \oiii and \nii and for the semi-strong lines \neiii, \ariii\
  and \hei. It contains 156 objects.

\end{enumerate}


\section{The observed N/O vs O/H diagram}
\label{noohobserved}


\subsection{Computation of temperature-based O and N abundances}

The abundances of O and N were recomputed in a homogeneous way using
5-level atoms for \Op, \Opp and \Np. The sources for the collision
strengths and transition probabilities are the following\footnote{Note
  that the atomic data used to compute the O and N abundances are the
  same as the ones entering in the version of \cloudy used to compute
  our grid of models.}. For O\,{\sc ii}: \citet{Kisielius.etal.2009a}
and \citet{Zeippen.1982a}; for O\,{\sc iii}:
\citet{Aggarwal.Keenan.1999a}, \citet*{Galavis.Mendoza.Zeippen.1997a},
and \citet{Storey.Zeippen.2000a}; for N\,{\sc ii}: \citet{Tayal.2011a}
and \citet{Galavis.Mendoza.Zeippen.1997a}. The electron densities were
computed from the \rSii\ ratio (when available) using the atomic data
from \citet{Tayal.Zatsarinny.2010a} and
\citet{Mendoza.Zeippen.1983a}. When the \rSii\ ratio was not available,
it was assumed that the electron density is equal to 100 cm$^{-3}$.

The ionic abundances were computed with a two-zone electron temperature
scheme. The temperature derived from \rOiii\ was used for \Opp\ and the
temperature derived from \rNii\ was used for \Op\ and \Np. When one of
the two line ratios was missing the following classical relation from
\citet{Garnett.1992a} was used:
\begin{equation}
\label{eq:T}
T_{\nii} = T_{\oii} = 0.70 \times T_{\oiii} + 3000\mathrm{ K}.
\end{equation}

The O and N abundances were obtained using the classical assumption that
oxygen is only in the form of \Op\ and \Opp\ in the \hii region and that
N/O $=$ \Np/\Op.

The uncertainties were estimated by a Monte-Carlo procedure using the
uncertainties on the observed line fluxes as described in detail in
\citet{Stasinska.etal.2013b}. Uncertainties due to possible deviations
from Eq.~\ref{eq:T} as well from the N/O $=$ \Np/\Op\ equation were not
taken into account in the Monte-Carlo procedure.


\subsection{Comparison of N/O vs O/H diagrams}

\begin{figure*}
\centering
\includegraphics[width=0.83\textwidth, trim=0 10 0 20]{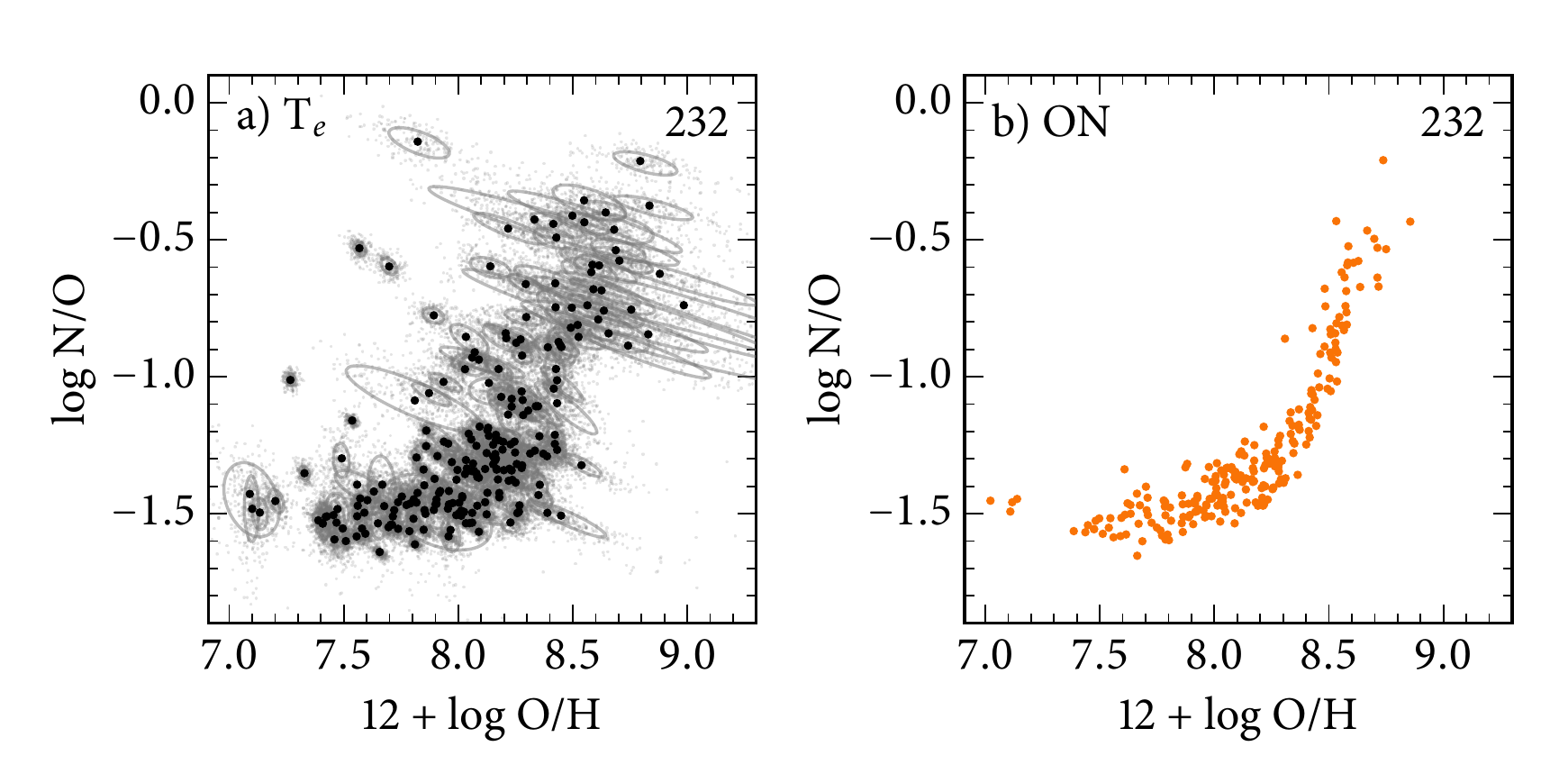}
\caption{The observed N/O vs O/H diagram for the objects of our sample T
  where \sii has been measured. (a) Diagram derived from the
  temperature-based method. The large dots represent the nominal
  solution, the small background dots the 200 Monte-Carlo realisations
  for each object, and the ellipses the covariances. (b) Diagram
  calculated by the ON method by \citet{Pilyugin.Vilchez.Thuan.2010a}
  for the same objects.}
\label{nooh-te-p10}
\end{figure*}

Fig.~\ref{nooh-te-p10} compares the N/O vs O/H diagram using two
different methods for the abundance determination. In the left panel we
used the temperature-based method as described above. In the right panel
we considered the ON method from \citet{Pilyugin.Vilchez.Thuan.2010a},
which is based on the \oii/\Hb, \oiiis/\Hb, \niis/\Hb, and \sii/\Hb
emission line ratios. Both panels show exactly the same 232 objects:
From our sample T we select only those objects where \sii has been
measured, which is necessary for the ON method.  This is a strong line
method calibrated on a sample of \hii regions with available
temperature-based abundances.  One can see that the two panels of
Fig.~\ref{nooh-te-p10} look very different, with the left one showing
significantly more dispersion than the right one.  Note that panel (a)
shows some points which are really far away from the main trend, while
their associated uncertainties are small.

Which of the two diagrams is closer to reality? Temperature-based
methods are often considered the most reliable. However this assertion
must be tempered by several considerations. Temperature-based methods
assume relations between some parameters (like $T_{\nii}$ and
$T_{\oiii}$ or N/O and \Np/\Op), whereas in fact some dispersion is
expected (see Appendix \ref{fake}). They are also strongly dependent on
errors in the intensities of the weak lines that serve to determine the
temperatures.  At the highest metallicities, important temperature
gradients inside the \hii regions may bias the abundance results, as
shown by \citet{Stasinska.2005a}. Shocks may contribute to the
intensities of the auroral lines, and falsify the results on
abundances. Finally, if the electron velocities in the ionized gas are
not Maxwellian but rather follow a $\kappa$ distribution as suggested by
\citet*{Nicholls.Dopita.Sutherland.2012a}, classical temperature-based
methods will result in underestimated abundances with respect to
hydrogen.  Because of all these reasons, it is not unreasonable to think
that part of the scatter observed in Fig.~\ref{nooh-te-p10} may be
artificial. On the other hand, the very tight relation between N/O and
O/H seen in Fig.~\ref{nooh-te-p10} right may be unreal, since the
formulae developed by \citet{Pilyugin.Vilchez.Thuan.2010a} tend to
strongly tighten any preexisting correlation (see Appendix
\ref{comparison}, Fig.~\ref{fig:fakeP10}).

In what follows, we develop a new method to derive oxygen and nitrogen
abundances in giant \hii regions which is much less affected by the
intensities of auroral lines than the temperature-based methods and,
unlike previous strong line methods, does not involve any assumption on
the N/O ratio.


\section{The model grid}
\label{modelgrid}


\subsection{Definition of the grid}
\label{griddef}

Because we want to avoid any biases in our method, we need to construct a
grid in which we vary all the determinant parameters. If we view a giant
\hii region as a nebula powered by an instantaneous burst of star
formation, the main parameters for our problem are the oxygen and
nitrogen abundances, the mean ionization parameter and the age of the
burst. The density distribution may also have a certain importance.

Using \cloudy 13.03 \citep{Ferland.etal.2013a}, we constructed a grid of
models defined as follows.

\begin{enumerate}

\item The oxygen abundance on the scale of 12 $+$ log O/H goes from 6.6
  to 9.4 in steps of 0.2 dex (15 values). The abundances of all the
  heavy elements except nitrogen and carbon are taken proportional to
  that of oxygen, as in \citet{Stasinska.etal.2015a}. The helium
  abundance varies with the oxygen abundances as in
  \citet{Stasinska.etal.2015a}.

\item The N/O ratio takes the logarithmic values
  $-2,\, -1.5,\, -1,\, -0.5,\, 0$. The abundance of carbon is linked to
  that of nitrogen by log C/H $= 0.48 + \log$ N/H.
  
\item Dust is included in the models, being related to the oxygen
  abundance in exactly the same way as in \citet{Stasinska.etal.2015a},
  following the works of \citet{RemyRuyer.etal.2014a} and
  \citet{Draine.2011c}.

\item The mean \textit{input} ionization parameter, defined by eq.~4 of
  \citet{Stasinska.etal.2015a}, takes the logarithmic values
  $-1,\, -1.5,\, -2,\, -2.5,\, -3,\, -3.5,\, -4$. Note that the real
  mean ionization parameter of the \textit{computed} model is somewhat
  different from the input value, since it depends on the electron
  temperature and on the partial absorption of the ionizing photons by
  dust \citep[see fig.~B2 of][]{Stasinska.etal.2015a}. In the remaining
  of the paper we denote this mean \textit{input} ionization parameter
  as $U$.

\item The starburst age takes the values of 1, 2, 3, 4, 5 and 6 Myr. The
  spectral energy distribution of the ionizing radiation is obtained
  from the population synthesis code PopStar
  \citep*{Molla.GarciaVargas.Bressan.2009a} for a \citet{Chabrier.2003a}
  stellar initial mass function and for the appropriate metallicity,
  obtained by interpolation.

\item In order to assess the effect that geometry might have, we
  consider two density distributions. One is a filled sphere of density
  $n = 100 \;\mathrm{cm}^{-3}$, the other is a thin spherical shell of same
  density. Roughly, the first scenario can correspond to a relatively
  young \hii region and the second to an evolved one. Clearly these
  choices are very simplistic and the role of the density distribution
  should be explored further. The mathematical definitions of the shell
  and filled sphere are detailed in section 4.1 of
  \citet{Stasinska.etal.2015a}.

\end{enumerate}

All the models are computed from the inner boundary until the ratio of
ionized hydrogen to total hydrogen density falls below 0.02.

\subsection{Some characteristics  of the grid}
\label{chargrid}

\defcitealias{Baldwin.Phillips.Terlevich.1981a}{BPT}

\begin{figure*}
\centering
\includegraphics[width=0.85\textwidth, trim=0 10 0 18, clip]{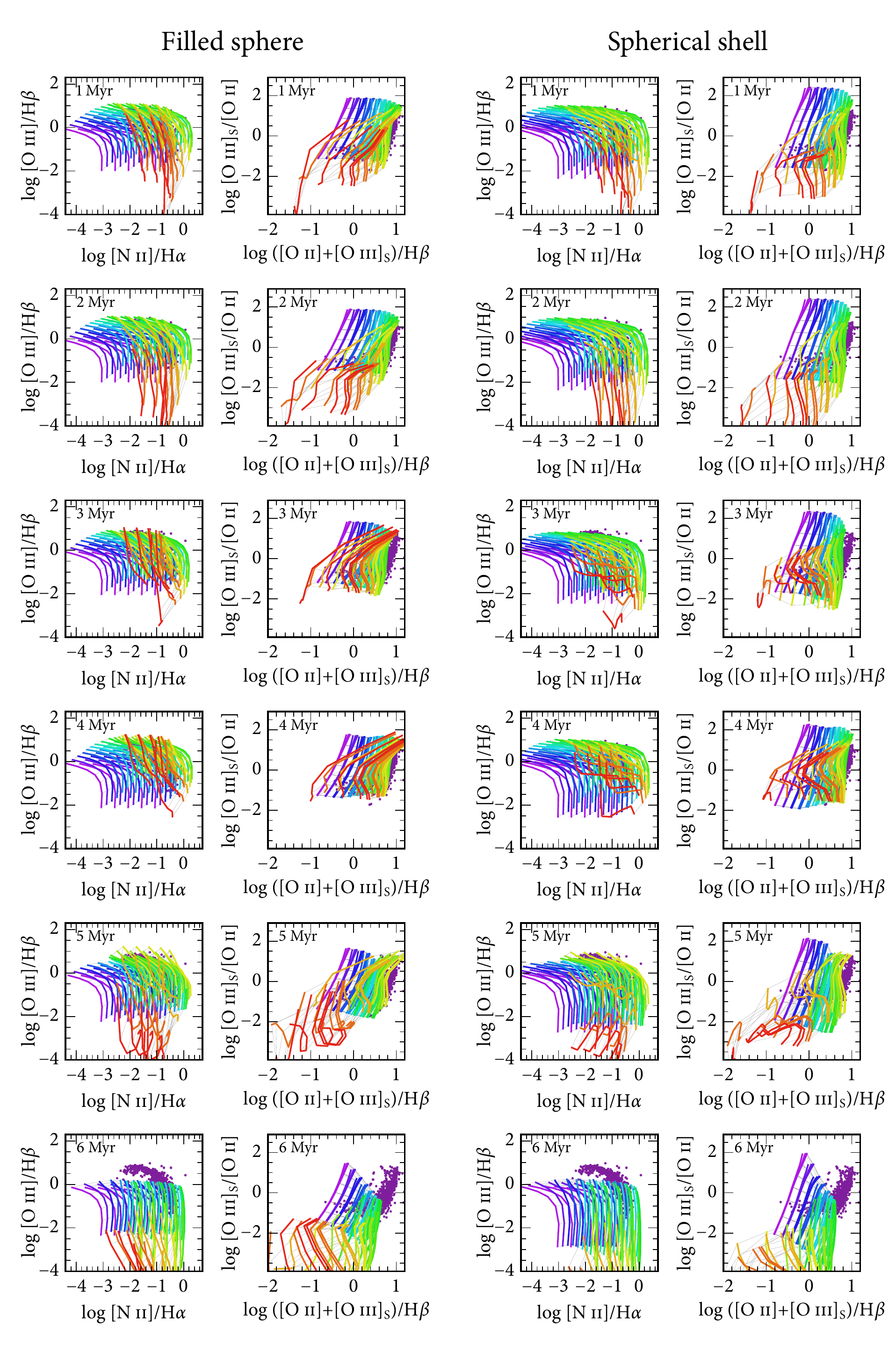}
\caption{The model grid on \oiii/\Hb\ vs \nii/\Ha
  \citepalias{Baldwin.Phillips.Terlevich.1981a}, and \oiiis/\oii vs
  (\oiiis+\oii)/\Hb diagrams for various ages (1 to 6 Myr) and
  geometries (filled sphere or empty spherical shell, indicated at the
  top of the figure). The aim of the postage-stamp size panels is
  twofold. First, one can see at a glance how the different parameters
  change the shape of the model subgrids. Second, the model grid is
  overplotted on the observational data, which allows one to judge how
  well the grid covers the observational points. The fact that most
  points are hidden behind the grid is thus a fact to be celebrated. The
  colour-coding of grid lines is detailed in Fig~\ref{fig:grid_zoom},
  which zooms into two panels. A colour version of this figure is
  available in the electronic edition. }
\label{fig:grid}
\end{figure*}

\begin{figure*}
\centering
\includegraphics[width=0.9\textwidth, trim=0 10 0 35, clip]{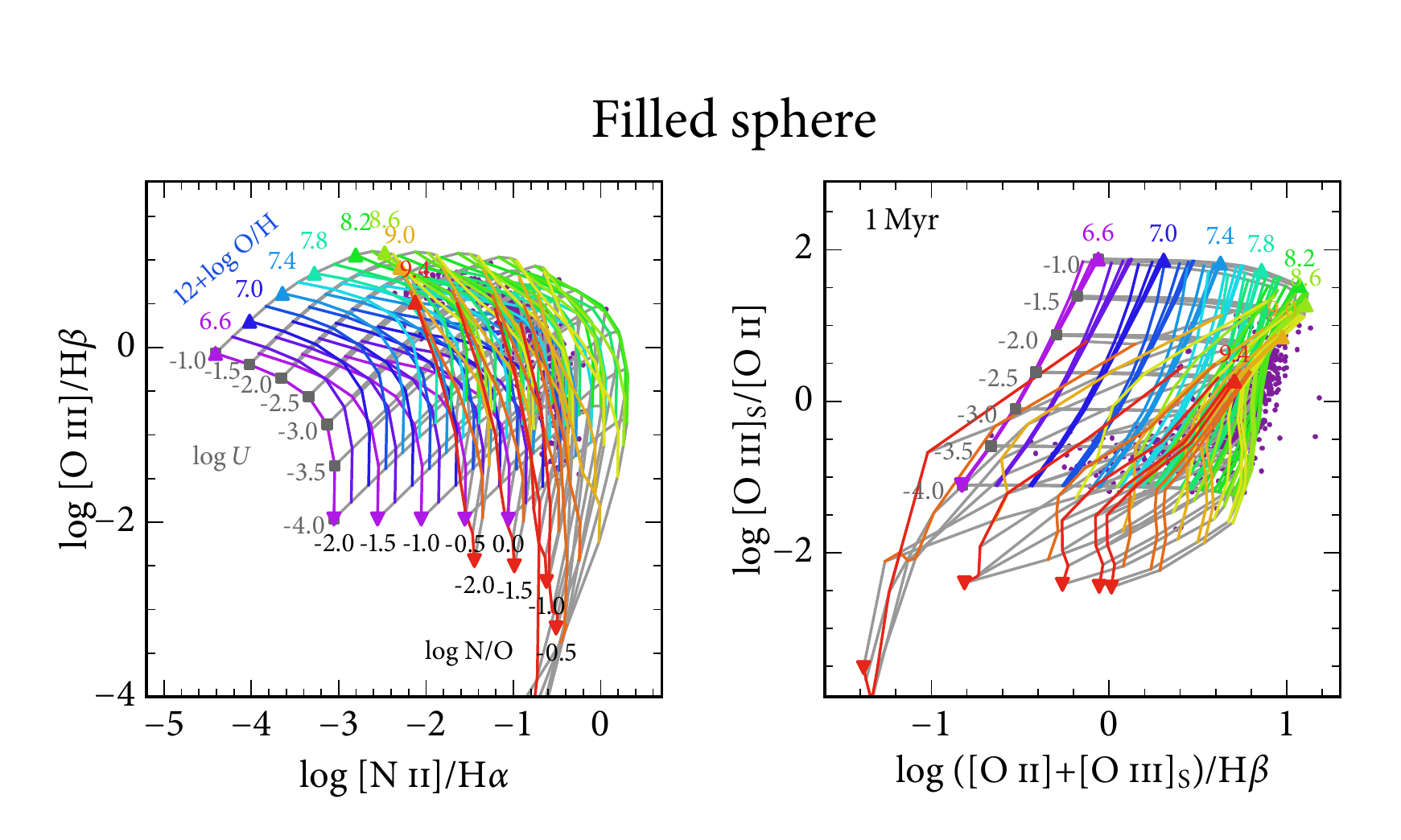}
\caption{Zoom into the top left panels of Fig.~\ref{fig:grid}. The lines
  mark the region covered by a subset of our grid of photoionization
  models (those with starburst ages of 1 Myr and a filled sphere nebular
  geometry). The small dots mark our observational data, which are
  mostly hidden behind the grid. Triangles pointing up, down, and
  squares mark a few values of O/H, N/O and $U$, respectively.  The
  colours of the grid lines change with the value of O/H, going from red
  for the largest values to purple for the lowest one, following the
  order of the rainbow.  Lines of the same colour have the same O/H but
  different N/O. Models with the same ionization parameter $U$ are
  linked by a grey line.  Readers interested in exploring our grid in
  detail can zoom in the pdf version of the paper or can download an
  electronic file with our grid of models from the BOND website
  (\url{http://bond.ufsc.br}) or the 3MdB database
  (\url{https://sites.google.com/site/mexicanmillionmodels/}). }
\label{fig:grid_zoom}
\end{figure*}

Fig.~\ref{fig:grid} shows the entire grid of computed models in two
planes frequently used to contemplate observations or models of \hii\
regions. One is the \oiii/\Hb\ vs \nii/\Ha\ plane often used for
excitation diagnostics (commonly called the BPT diagram after
\citealp*{Baldwin.Phillips.Terlevich.1981a}), and the other is
$\oiii \lambda 5007+4959 / \oii \lambda 3727$ vs
$(\oiii \lambda 5007+4959 + \oii \lambda 3727) / \Hb$ introduced by
\citet{McGaugh.1991a} to derive the oxygen abundance and to which we
will refer as the McG diagram. The models for filled spheres are
displayed in the left column, while the models for shells are displayed
in the right column. Each row of panels corresponds to a given starburst
age, increasing downwards.  Fig.~\ref{fig:grid_zoom} is a zoom in the
BPT and McG planes for the 1 Myr filled sphere subgrid, which serves to
better illustrate the coloured lines and points drawn in
Fig.~\ref{fig:grid}.  Models joined with full coloured lines have the
same O/H and same N/O (the colour is defined by the value of O/H and
runs from purple to red following the rainbow colours as O/H increases),
while models joined with thin grey lines have the same input value of
the mean ionization parameter. In all the panels the model curves are
superimposed on the observational points.

The first thing we can notice is that the entire grid appears to cover
most of the observational points in these two planes, which is what we
were looking for, i.e. the fact that the observational points are
difficult to see in the figure is a feature, and not a flaw.  However,
in the McG diagram, a small proportion of objects appears slightly to
the right of the grid, at any of the ages considered, meaning that in
this region the electron temperature computed by the models is probably
lower than in real \hii regions. We have explored several possibilities
to reduce the problem by playing with dust and abundance ratios and
density, but we did not succeed. Anyway, the discrepancy is much smaller
than in the studies of \citet{McGaugh.1994a} and \citet[especially for
their grid with a $\kappa$ distribution of
electrons]{Dopita.etal.2013a}. We think that the discrepancy we find is
due to our models still not reproducing exactly real objects rather than
to observational errors.  Nevertheless, the magnitude of the problem is
sufficiently small to warrant our further use of the grid for abundance
determinations.  Indeed, we find that excluding objects that fall off
grid do not change our results.

The two different density distributions (filled sphere and empty shell)
produce only slight apparent differences in the grid but, as we will see
in Appendix~\ref{tests}, this is sufficient to affect the O/H and N/O
ratios by up to 0.05 dex. More realistic density distributions, such as
a core-halo density distribution or a constant pressure distribution may
have a larger impact.

The variations in N/O obviously have an impact on the
\citetalias{Baldwin.Phillips.Terlevich.1981a} diagram but they also
affect the McG diagram at high metallicities, since they affect the
cooling rates. In other words, if the N/O ratio is abnormally high, this
would bias the O/H derived from strong-line methods not accounting for a
possible scatter in N/O.

We also note that at the highest metallicities the curves of equal
chemical composition become ill-behaved. This is because, at such
metallicities, most of the cooling occurs through the infrared
fine-structure lines whose intensity is not very sensitive to
temperature. Therefore a small change in the physical conditions of the
gas may alter the electron temperature considerably, which, in turn,
strongly affects the intensity of the \Oiii line. This means that, for
$12 + \log \mathrm{O/H}$ greater than, say, 9.2, the real error in the
abundances derived from optical lines is probably larger than can be
estimated from our grid.

Fig.~\ref{fig:grid} shows that the starburst age modifies the shape of
the model grid, especially in the McG diagram, so that assuming the same
age for all the \hii regions will produce significant errors in the
oxygen abundance determination. What actually changes from one age to
another is the `hardness' of the ionizing radiation field, i.e.\ its
capacity of heating the surrounding medium by photoionization.

\begin{figure}
\centering
\includegraphics[width=0.8\columnwidth, trim=0 10 0 10, clip]{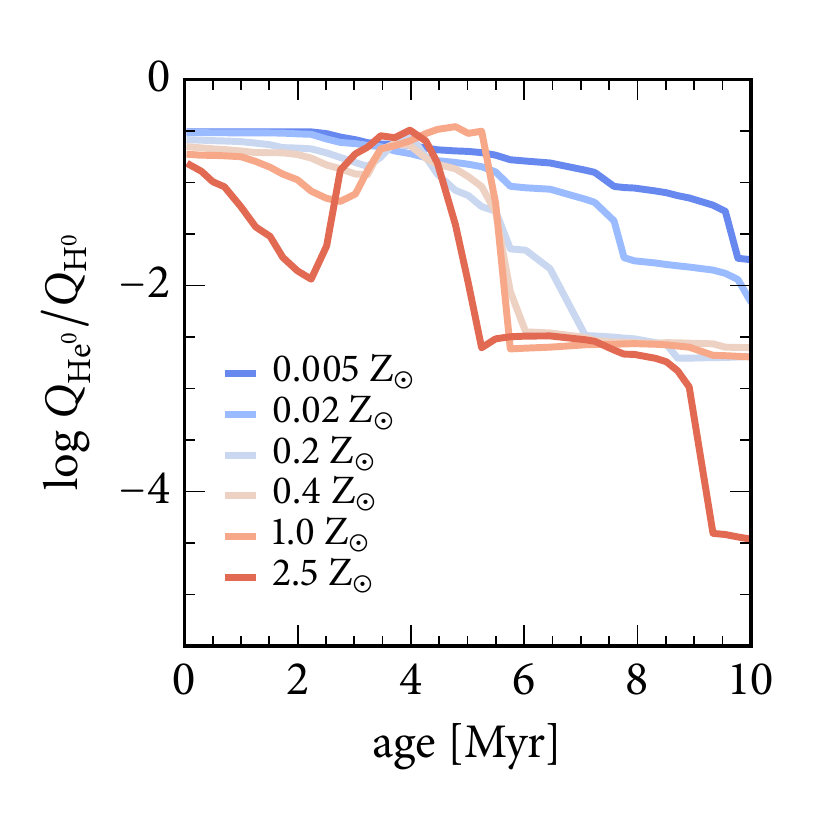}
\caption{The values of \qhe/\qh as a function of time for the PopStar
  models at different metallicities. The hardness of the radiation field
  in an \hii region tends to decrease with age, but the Wolf-Rayet
  phases wreaks havoc on this simple view.}
\label{fig:qheqh}
\end{figure}

The hardness can be viewed as the ratio \qhe/\qh, where \qhe is the
number of photons above 24.6 eV and \qh is the number of photons above
13.6 eV.  Fig.~\ref{fig:qheqh} shows the variations of \qhe/\qh as a
function of time for the six PopStar metallicities. Generally, the
ionizing radiation field softens as metallicity increases. However,
during the Wolf-Rayet phase the radiation field hardens and this effect
is higher at high metallicity. As a result, the radiation field is the
hardest at the highest metallicities and at ages around 3--5 Myr. This
implies that for these ages the $(\oiiis+\oii)/\Hb$ ratio can reach quite
high values at high metallicities.

In reality, the process of star formation may not be instantaneous, as
in the PopStar models, but extend over a certain time.  In practice,
what is important for the line intensities is not so much the age of the
ionizing stellar population or the regime of star-formation, but rather
the hardness of the resulting ionizing radiation field. Our models with
different ages should thus be viewed as models for spectral energy
distributions of different hardness.

\begin{figure}
\centering
\includegraphics[width=0.99\columnwidth, trim=0 10 10 7, clip]{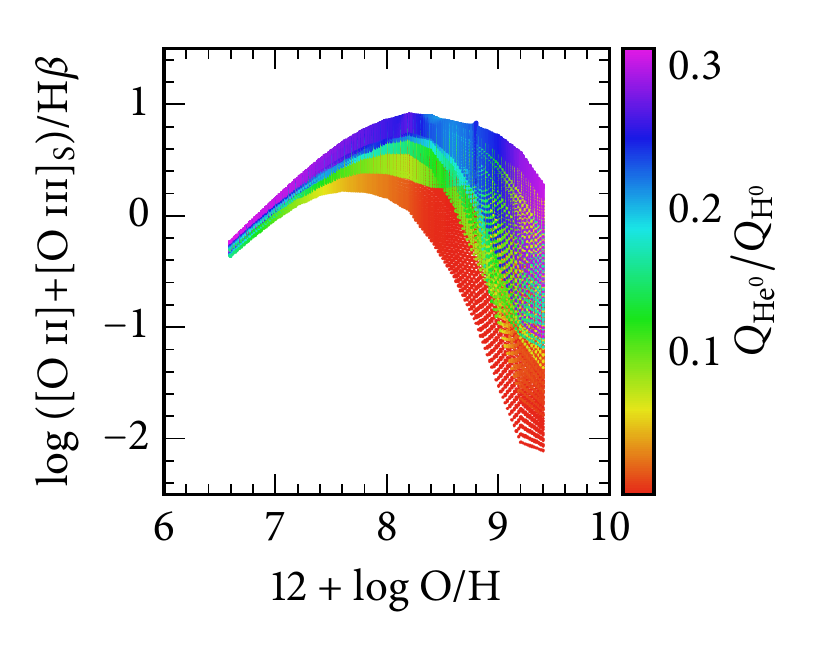}
\caption{$(\oiiis+\oii)/\Hb$ vs O/H for $\log U = -2$ and all the values
  of N/O, ages and density structures in our grid coloured by \qhe/\qh.
  In order to see the trends better, we have created a very fine
  interpolated grid for this figure (sampled 0.02, 0.05, 0.05 dex in
  O/H, N/O and $U$ respectively).  Note that for high metallicities the
  $(\oiiis+\oii)/\Hb$ ratio does not pin down the value of the O/H
  abundance. A colour
    version of this figure is available in the electronic edition.}
\label{fig:oqheqh}
\end{figure}

Fig.~\ref{fig:oqheqh} shows the variations of $(\oiiis+\oii)/\Hb$ with
respect to O/H for models with $\log U = -2$, the curves representing
the sequences of models being coloured according to the value of
\qhe/\qh. It is clear that for $12 + \log \mathrm{O/H}$ larger than, say,
7.5, the primary metallicity indicator, $(\oiiis+\oii)/\Hb$, is strongly
dependent on \qhe/\qh, reinforcing the relevance of considering this
effect in the modelling.


\section{Description of the BOND method}
\label{method}

Our grid of models spans a wide range of physical parameters: N/O, O/H,
$U$, the hardness of the radiation sources, and the density profile of
the nebula. Although we are only interested in inferring the nitrogen
and oxygen abundances, we need to constrain the other nuisance
parameters. This section explains our choices of observational
constraints and the formalism for our Bayesian Oxygen and Nitrogen
abundance Determinations (\bond) method.

\subsection{Observational constraints}
\label{constraints}

\subsubsection{Uncovering N/O and O/H}
\label{constraints-SL}

Our first set of observational constraints are the extinction-corrected
line ratios $\log \, \nii/\Hb$, $\log \, \oii/\Hb$ and $\log \, \oiii/\Hb$. The
formal assumption is that those logarithmic line ratios are Gaussianly
distributed and independent. Using line luminosities instead of line
ratios is meaningless for our models, since the models are not defined
by luminosities but by ionization parameters.

The physical reasoning behind using this set of line ratios is that
$\nii/\oii$, $(\oiii+\oii)/\Hb$, and $\oiii/\oii$ are proxies for N/O, O/H
and $U$, respectively.  Since we are using only strong lines, there is
no constraint on the electron temperature. That way, because
$(\oiii+\oii)/\Hb$ vs O/H is bivalued, this first set of constraints
finds bivalued solutions for O/H.

We believe that using carefully chosen emission lines is better than
using all information available for this problem. We therefore choose
not to include \sii for two reasons. \sii comes from the outskirts of
the nebula, which do not coincide with the region where the other strong
lines are produced. Any density structure will change \sii in relation
to the other lines. Second, the S/O ratio in \hii\ regions could be
subject to variations due to different production sites of S and O
(e.g.\ it has recently be proposed by
\citealp{DelgadoInglada.etal.2015a} that intermediate-mass stars could
contribute to the global oxygen budget in galaxies) and to different
dust-depletion schemes.

\subsubsection{Eliminating the bimodality}
\label{bimod}

\begin{figure}
\centering
\includegraphics[width=0.84\columnwidth, trim=0 10 0 10, clip]{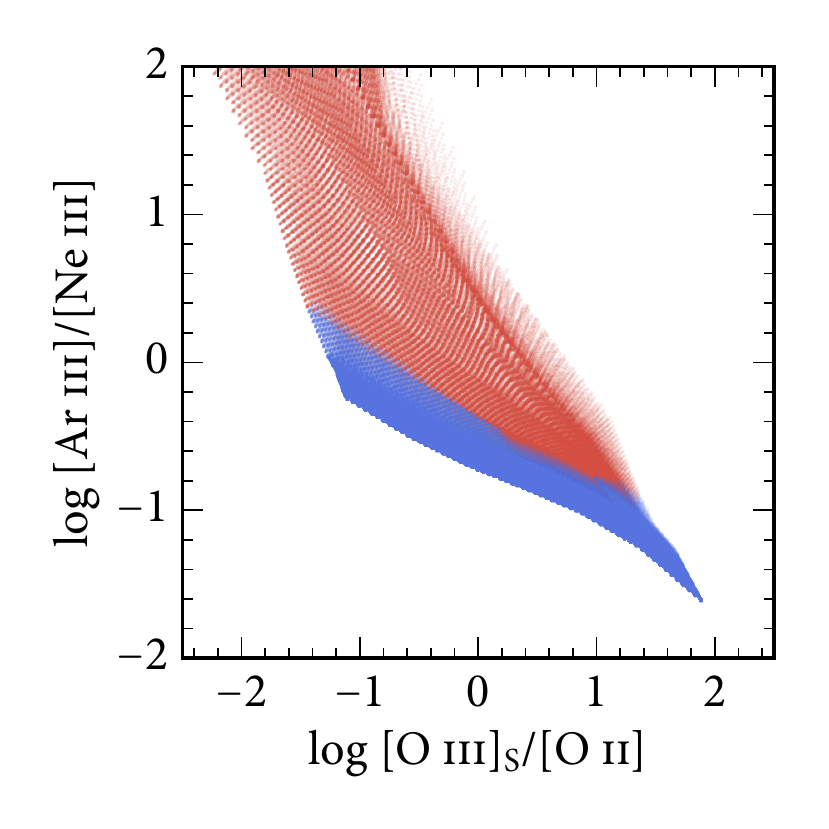}
\caption{\Ariii/\Neiii vs \oiiis/\oii for age 2 Myr and filled sphere of
  our grid (the same used in Fig.~\ref{fig:oqheqh}). For each
  value of N/O and $U$, we tag a model as being in the high or low-O/H
  branch from the $(\oiii+\oii)/\Hb$ vs O/H diagram.  We colour the low
  and high branches in blue and red, respectively.  Note how the two
  branches are almost cleanly separated in this diagram, with very
  little overlap of blue and red points. A colour
    version of this figure is available in the electronic edition.}
\label{fig:branches}
\end{figure}

As explained in the introduction, using nitrogen to break the
$(\oiii+\oii)/\Hb$ degeneracy with O/H is not satisfactory since the
relation between the N/O ratio and O/H is likely dispersed. It is better
to use a physical argument that does not depend on astrophysical
conditions, like one based on the electron temperature, which will be
low in the high abundance regime and high in the low abundance one. 

We need a line ratio that is easy to observe, and that depends strongly
on the electron temperature and weakly on ionization conditions and
abundance ratio. The \ariii/\neiii ratio fulfils these requirements.
The \ariii and \neiii lines have different excitation thresholds (1.7
and 3.2 eV, respectively), so different dependencies on the electron
temperature. Argon and neon are two primary elements; in addition, they
are both rare gases not suspected of dust depletion, so their abundance
ratio is expected to be constant. The \ariii and \neiii lines
do not arise exactly from the same zone, the ionization potential of
\Nepp\ being higher than that of \Arpp, but we can use the \oiii/\oii
ratio to figure out what the ionization is.

Fig.~\ref{fig:branches} shows \ariii/\neiii as a function of \oiii/\oii
for our subsample of models corresponding to an age of 2 Myr and a
filled sphere nebula.  The colours indicate the abundance regime: Blue
corresponds to the low metallicity branch in the $(\oiii+\oii)/\Hb$ vs
O/H diagram, and red to the high metallicity branch. We can see how
using the \ariii/\neiii ratio in conjunction with the \oiii/\oii one
separates the two branches.  In practice, we use $\log \, \ariii/\Hb$ and
$\log \, \neiii/\Hb$ separately as our constraints because they are closer
to an ideal Gaussian distribution than \ariii/\neiii.  Also, because the
grid was not designed to closely explain \ariii/\neiii, we add an extra
noise $e$ in quadrature to \ariii/\Hb and \neiii/\Hb and marginalise it
away, as explained in Section~\ref{marg-uncertain}.

The \ariii/\neiii as a function of \oiii/\oii for all starburst ages and
scenarios is a bit fuzzier, showing some superposition of high and low
metallicity grid points. For those cases, it is useful to include an
extra constraint to exclude the grid points on the wrong O/H branch.  We
find that using the \Oiiit and \Niit upper limits as additional
criteria for the electron temperature improves our solutions.  The
reasoning for using upper limits is that, if the \Oiiit (or \Niit) line
is not observed, and the observational upper limit for its intensity is
below the expected value in the low metallicity regime, this implies
that we are in the high metallicity regime. We include those upper
limits in our inference as discussed in Section~\ref{weak-lines}.  We
check that this does not force our solutions to match the observed
\Oiiit or \Niit.

\subsubsection{Characterising the radiation field}
\label{hardness}

\begin{figure}
\centering
\includegraphics[width=0.99\columnwidth, trim=0 10 0 10, clip]{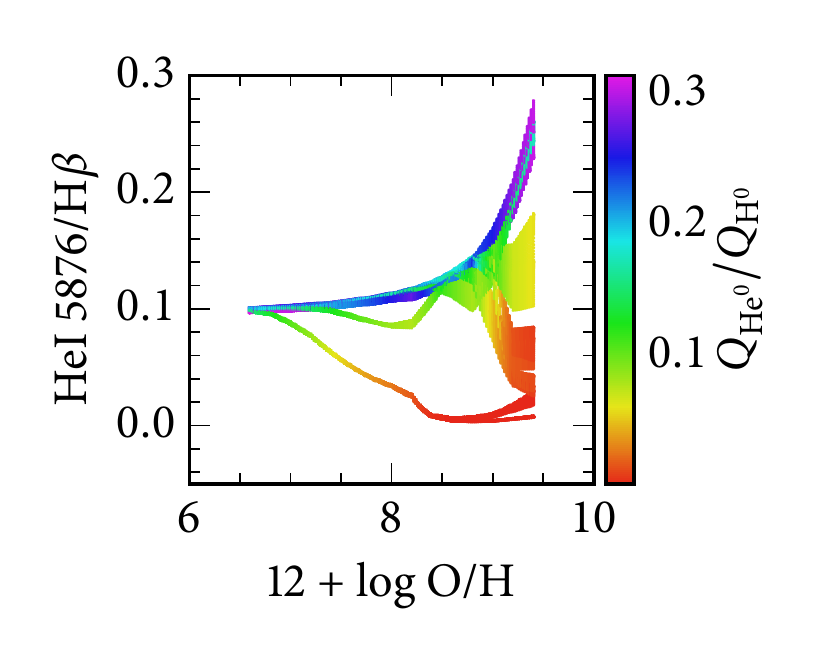}
\caption{O/H vs \Hei/\Hb coloured by \qhe/\qh for $\log U = -2$ and all
  the values of N/O, ages and density structures in our grid
  (the same used in Fig.~\ref{fig:oqheqh}).  For a given O/H, \Hei/\Hb
  is a good proxy of \qhe/\qh.  For the highest \qhe/\qh, however,
  \Hei/\Hb does not change much (blue and purple points all fall on the
  same region).  The sparsity of the stellar ages in our grid shows up
  as the large uncovered parts of this diagram. A colour
    version of this figure is available in the electronic edition.}
\label{heishb}
\end{figure}

Line ratios such as (\oiii/\oii)/(\siii/\sii)
\citep{Vilchez.Pagel.1988a} or (\ariv/\ariii)/(\oiii/\oii)
\citep{Stasinska.etal.2015a} can be used to estimate the mean effective
temperature of the radiation field. However, \ariv is often too weak to
be measured in giant \hii\ regions. (\oiii/\oii)/(\siii/\sii) does
depend on $U$ \citep[see][]{Stasinska.etal.2015a}, but, because it
depends on \sii, it suffers from the problem that it is affected by the
density distribution in real \hii regions mentioned above
(Section~\ref{constraints-SL}).

Another potential indicator is \Hei/\Hb which, as long as helium is not
fully ionized in the \hii region, is dependent on the spectral energy
distribution of the ionizing stars. The \hei/\Hb ratio, however, depends
on the metallicity, since the dependence of the \hei and \Hb line
emissivities with electron temperature is not the
same. Fig.~\ref{heishb} shows \hei/\Hb as a function of O/H in our grid
of models, coloured according to \qhe/\qh with the same colour scale as
in Fig.~\ref{fig:oqheqh}. We see that, with the information on O/H given
by the other lines used in \bond, \hei/\Hb allows one to estimate
\qhe/\qh\ up to a value of $\sim$ 0.2. This is not entirely
satisfactory, because Fig.~\ref{fig:oqheqh} shows that at the highest
metallicities, the value of (\oiiis+\oii)/\Hb depends on \qhe/\qh also
at values larger than 0.2.  In this paper we restrict ourselves to using
\hei/\Hb to characterise the hardness of the radiation field. In future
works, when large data bases of giant \hii regions with fully described
deep spectra become available, it will be possible to add information on
\ariv/\ariii.  We also allow an extra noise $e$ and integrate it out for
$\log \hei/\Hb$ as described in Section~\ref{marg-uncertain}.


\subsection{The probabilistic formalism}
\label{bayes}

We aim to find the oxygen and nitrogen abundances of an \hii region by
comparing its observed lines $O$ to our grid of models. A given
observed line is characterised by its intensity and uncertainty
($o_j, \sigma_j$), and an \hii region by its $j=1..J$ emission lines:
\begin{equation}
\label{eq:obs}
O \equiv \{ O_j \} = \{ o_j, \sigma_j \},
\end{equation}
where the curly braces define a set of values spanning the rightmost
index (i.e.\ $j$ for the equation above).  Each model $M$ in our grid
is defined by its $i=1..I$ model parameters $m_i$, and 
generates a set of computed line intensities $\{c_{j}\}$:
\begin{equation}
\label{eq:model}
M \equiv \{ \{m_i\}; \{c_{j}\} \}.
\end{equation}

Our grid of models spans a wide range of values not only for our two
parameters of interest (oxygen and nitrogen abundances), but also for
the ionization parameter, starburst ages and nebular geometry.  While
the latter play an important role in the photoionization modelling of an
\hii region, we do not wish to infer them.  From a pragmatic point of
view, a Bayesian formalism offers a framework to {\it marginalise away}
those nuisance parameters by simply integrating them out. This comes
with the cost of writing down the {\it posterior} probability so that
the dimensions of our probability function still make physical sense
after those integrals are performed (see e.g.\ \citealp{Hogg.2012a}).
The posterior probability density function (PDF) for a model $M$ given
the observed data $O$ and any other relevant background information $B$
is
\begin{equation}
\label{eq:bayes}
p(M | O, B) = {\cal N} \, p(M | B) \; p(O | M, B),
\end{equation}
where PDFs are written as $p$, and ${\cal N}$ is a normalisation
constant so that the posterior integrates to unity over all the
parameter space. The PDFs on the right hand side of the equation are the
prior probability of the model parameters, and the likelihood of
observing $O$ assuming $M$ and $B$ are true. In what follows we will
discuss our generative model for the likelihood and our choice of the
prior.


\subsection{Generative model for data points}
\label{likelihood}

\subsubsection{Partially marginalised likelihoods}
\label{marg-uncertain}

We assume Gaussian uncertainties for our constraints, which are the
logarithmic line fluxes with respect to \Hb. This is a good
approximation given that we consider high S/N observations, so that \Hb
is very well determined and the line ratios noise uncertainties must
deviate very little from a Gaussian distribution.  The likelihood of
observing an emission line $O_j = ( o_j, \sigma_j )$ given the model $M$
and the background information $B$, plus an extra source of noise with
dispersion $e$ is
\begin{equation}
\label{eq:likel}
p(O_j | e, M, B) = {\cal N^\prime} \frac{1}{\sqrt{\sigma_j^2 + e^2}}
                     \exp{\left[ -\frac{(c_{j} - o_j)^2}{2(\sigma_j^2 + e^2)} \right]} ,
\end{equation}
where ${\cal N^\prime}$ is a normalisation constant.

\begin{figure}
\centering
\includegraphics[width=0.8\columnwidth, trim=0 15 0 10, clip]{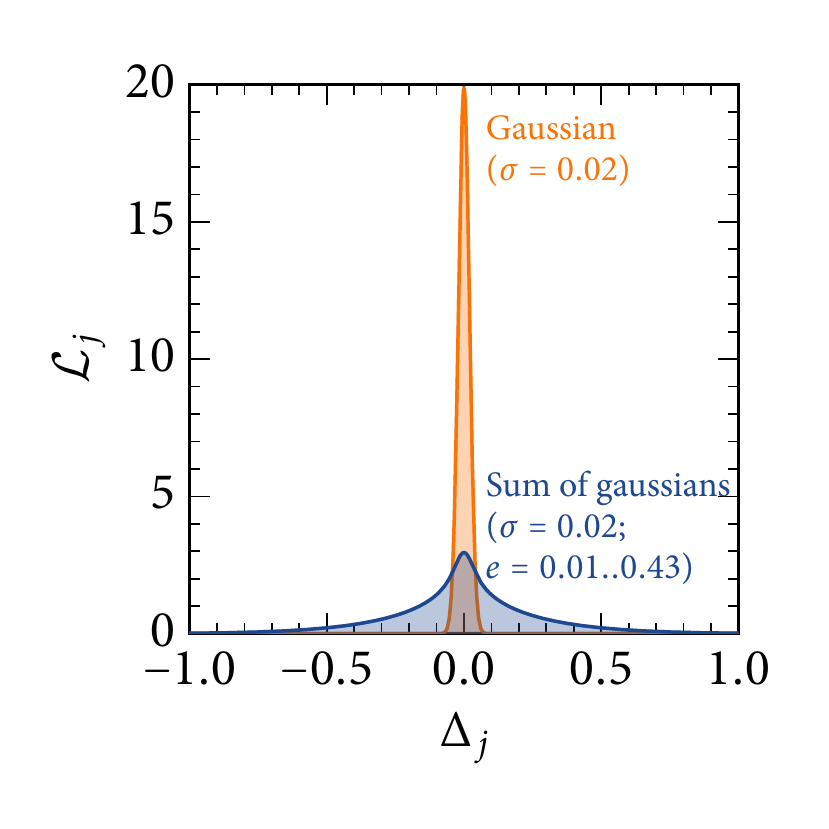}
\caption{Comparison between a Gaussian, in orange, centred in
  $\Delta_{j}$ and dispersion $\sigma_j = 0.02$ dex (the
  typical uncertainty of semi-strong line ratios $\ariii/\Hb$,
  $\neiii/\Hb$ and $\log \Hei/\Hb$ in our sample B), and a sum of
  Gaussians, in blue, with dispersions $(\sigma_j^2 + e^2)^{1/2}$ for
  $e$ between 0.01 and 0.43 dex (corresponding to 2 to 100 per cent of
  the measured line intensity). Both functions are scaled so that their
  areas are equal to unity. }
\label{fig:marg_uncert}
\end{figure}

The term $e$ was introduced in order to account for deviations in
emission line ratios not contemplated in our models. For instance, we
use \ariii/\Hb and \neiii/\Hb as constraints, but our models were not
meant to reproduce the argon and neon abundance dispersions in nature.
For constraints involving strong lines (\nii, \oii, \oiii) we simply set
$e = 0$.  For constraints based on semi-strong lines (\ariii, \neiii,
\hei), we consider $e$ in a interval of 0.01 to 0.43 dex (i.e.\ for an
extra noise from 2 to 100 per cent of the measured line
intensity). Since we are not interested in $e$, we integrate it out to
calculate the marginalised likelihood for $O_j$:
\begin{equation}
{\cal L}_{j} = p(O_j | M, B) = {\cal N^{\prime\prime}} \int p(e |
M, B) \; p(O_j | e, M, B) \, de,
\end{equation}
where ${\cal N^{\prime\prime}}$ is yet another normalisation constant.
In practice we calculate ${\cal L}_{j}$ as a sum of Gaussians for
logarithmically spaced values of $e$. Using only 20 Gaussians for this
sum guarantees that the numerical integral with $\log$-spaced $e$ is
equivalent down to $\sim 10^{-4}$ to a numerical integral with linear
spacing of $10^{-6}$ dex.

Fig.~\ref{fig:marg_uncert} compares a Gaussian with $\sigma_j = 0.02$
dex (a typical value for the semi-strong lines in our sample) centred in
$\Delta_{j} = c_{j} - o_j$ and $e = 0$ to a sum of Gaussians with
variances $\sigma_j^2 + e^2$, with $e$ varying from 0.01 to 0.43
dex. The marginalised likelihood for the sum of Gaussians is much
broader than the likelihood for a single Gaussian, but note that it
eventually drops to zero. This is an effect brought about by the term
$(\sigma_j^2 + e^2)^{-1/2}$ in Eq.~\ref{eq:likel}, which penalises very
large values of $e$. Therefore, although we allow $e$ to vary, the
likelihood is still shaped by the data. What happens in practice is that
we probe regions of the emission line space in our model grid which are
far from the nominal observed measurement.

\subsubsection{Treating weak lines}
\label{weak-lines}

We do not constrain weak line intensities such as \Oiiit or \Niit, but
we use their upper limits as an additional temperature constraint.  The
upper limit $u_j$ of the weak line $j$ is defined as the 2-$\sigma$
detection limit for a given spectrum. The likelihood for weak lines is a
step function that masks out models whose computed emission lines $c_j$
are above the upper limit $u_j$:
\begin{equation}
p(O_j | M, B) \propto
\left\{
  \begin{array}{ll}
    1 & \mbox{if } c_{j} \leq u_j, \\
    0 & \mbox{if } c_{j}   >  u_j.
  \end{array}
\right.
\end{equation}

This constraint is useful when the \ariii/\neiii ratio alone is unable
to distinguish the low and high temperature branches. Imposing an upper
limit on \Oiiit or \Niit flags out the solutions for which the
temperatures are too high.

This constraint is peculiar, since it is only available for undetected
\Oiiit or \Niit lines. Increasing the quality of the observations means
that \Oiiit and/or \Niit would be detected and we would not be able to
use this constraint any longer. In order to avoid the asymmetry of
having this constraint applied for some objects and not others, we
impose our upper limit criteria even for sources where \Oiiit and/or
\Niit are detected. This procedure also guarantees that higher S/N data
are not penalised. We thus assume that the upper limit for a detected
line is its intensity plus its 2-$\sigma$ uncertainty, which again helps
clear out solutions with too high a temperature.

\subsubsection{Taking  all constraints into account}
\label{likel-constraints}

Assuming that the observed line intensities $O_j$ are independent, the
likelihood for all observed line intensities $O$ for a given model $M$ is
\begin{equation}
p(O | M, B) = \prod_j p(O_j | M, B).
\end{equation}

We usually write this down as
$\ln p(O | M, B) = \sum_j \ln p(O_j | M, B)$.  Expressing the likelihood
in natural logarithmic highlights two points. First, we emphasise that
our code adds up instead of multiplying values to minimise numerical
errors. Second, we see that, in the case of a fixed extra noise source
(constant $e$), the likelihood reduces to
$\ln [-0.5 \sum_j {(c_{j} - o_j)^2}/{(\sigma_j^2 + e^2)]} \equiv -0.5
\chi^2$
apart from a constant of proportionality.  We warn however that the
familiar $\chi^2$ minimisation should be looked with suspicion when
applied to abundance determinations. First, when strong and weak lines
are all fitted at the same time, weak lines are penalised for having
larger uncertainties. Albeit formally correct, this lessens the
importance of weak lines while they may carry important information, for
example, \Oiiit or \Niit, which pin down the electron temperature.
Second, the $\chi^2$ can compensate one badly fitted line with one that
is extremely well fitted. The correct way to fit photoionization models
would be to fit \textit{each} line within an appropriate error-bar,
which is not ensured by calculating the likelihood by the $\chi^2$.  To
some extent our method is immune to this problem because we only use
strong and semi-strong lines as constraints that strongly shape the
likelihood, while weak lines are only used as upper limit measurements.


\subsection{Adaptive octree grids}
\label{prior-octree}

The missing piece to calculate the posterior is the prior PDF. Our
background knowledge $B$ (hence our prior) is encoded in the sampling of
our model grid.  We follow the reasoning by \citet{Blanc.etal.2015a} and
assume a flat logarithmic prior on O/H, N/O and $U$. This is equivalent
to a Jeffrey's prior for a Gaussian distribution with fixed standard
deviation. The age of the ionizing source is linearly sampled, and we
have two nebular geometries (a filled sphere and a spherical shell); we
assume a flat prior for those.

Our original grid is finely meshed in O/H (0.2 dex), but coarse in all
other parameters (0.5 dex in $U$ and N/O, 6 starburst ages from 1 to 6
Myr and two nebular geometries). The emission line space is consequently
sparsely sampled. When uncertainties in the data are much smaller than
the distance between grid models, very few models will be near the
observed data. Creating a finer grid can mitigate this problem. Running
a sufficient number of photoionization models to fill in the ionizing
source ages and nebula density structures adequately would however be
unnecessarily time-expensive. Interpolating our grid solves the grid
sparsity problem quickly and is a good approximation, since the emission
line intensities vary smoothly with those parameters once the initial
grid is dense enough. We thus interpolate our original grid in
$\log \mathrm{O/H}$, $\log U$ and $\log {N/O}$, but not in starburst age
and geometries, which would be dangerous and meaningless,
respectively. A finer grid in the latter parameters would require
running more photoionization models, a time-consuming task both for
generating the grid and running the \bond code. We choose to keep ages
and geometries fixed, which pop up as discontinuities and multimodal
solutions in our posterior PDFs (take a sneak peek at e.g. the `islands'
of solutions in Fig.~\ref{fig:marg_pdfs}).

\begin{figure}
\centering
\includegraphics[width=0.8\columnwidth, trim=0 0 0 30, clip]{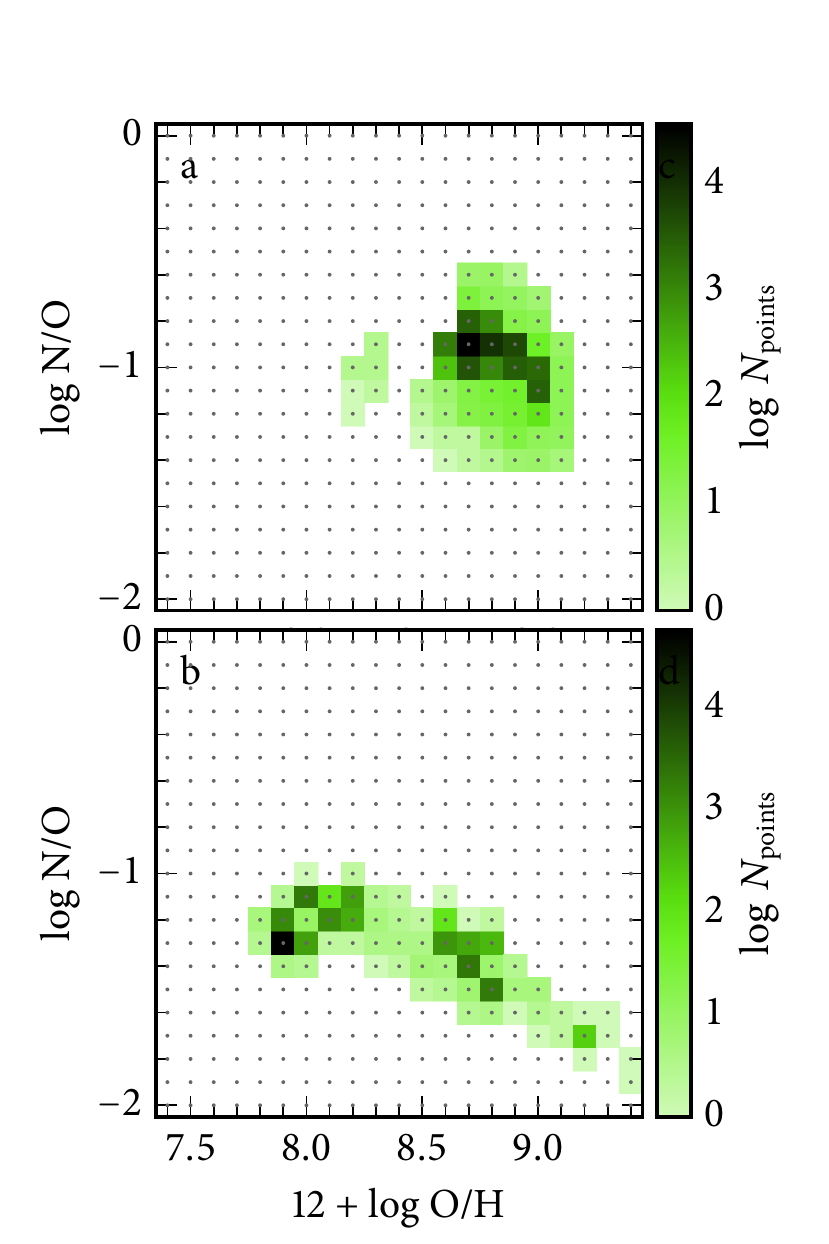}
\caption{Original grid cell centres (dots) and the final number of
  octree grid cells (green scale) for two sources in our sample.  Panel
  a is for an \hii region in NGC 1232 (slit number 5 from
  \citealp{Bresolin.etal.2005a}) and panel b for the blue compact dwarf
  HS0837+4717.  }
\label{fig:octree}
\end{figure}

We create a different interpolated grid adapted to each object using an
octree sampling algorithm. Octree grids are usually applied to sample a
Cartesian 3D space, and are used extensively in video games, computer
graphics, hydrodynamics simulations, and Monte Carlo radiative transfer
codes (e.g.\ \citealp{Saftly.etal.2013a} and references therein).

We start off with a grid containing $226,548$ models separated by 0.1
dex in O/H, $U$ and N/O. Starburst ages and nebular geometries are kept
fixed.  Each grid point represents a cell of volume
$dV = d(\log \mathrm{O/H}) \, d(\log \mathrm{N/O}) \, d(\log U)$. For each
object, we calculate the posterior PDF for all grid points, and the
contribution $dP = p(M | O, B) \, dV$ of each grid cell to the total
probability.

After this first run, we remove grid cells which contribute too little
to $P$ to speed up the calculations (the default option is to remove
grid points for which $dP < 10^{-20}$ considering each age and geometry
scenario separately). Grid cells where $dP \ge 10^{-4}$ are subdivided
into eight subcells, with each subcell corresponding half the size of
the parent cell in O/H, $U$ and N/O. The $10^{-4}$ threshold was chosen
as a compromise between the precision of the posterior PDF and the
computing time.  Smaller thresholds create very large octree grids,
whilst our nominal solutions (i.e. the posterior summaries, see
Section~\ref{posterior}) for $\log \mathrm{O/H}$, $\log \mathrm{N/O}$ change
by less than 0.02 dex.  We recalculate the posterior with the new octree
grid, subdivide the cells where needed, and reiterate until there is no
remaining cell with $dP$ above the threshold.  This procedure creates a
grid that is finer in the parameter space region where the posterior
probability is higher.

Fig.~\ref{fig:octree} shows the final octree grid for two objects in
sample B, compressed in the N/O vs O/H space. The dots are the centres
of the original 0.1 dex-sampled grid cells, and the green scale
represents the final number of subcells.  Swathes of white space stand
for grid cells that have been removed due to contributing too little to
the final probability.  For our sample B, the median number of cells is
$\sim65,000$ (for a minimum and a maximum of $31,604$ and $79,555$), and
all the grid cells usually go below the $dP$ threshold after 6
iterations, which yields subcells $8^{-6} = 1/262,144$ times smaller
than the original cell (i.e.\ which span 0.0015625 dex in O/H, N/O and
$U$).  The average time to run \bond for one source with the octree
sampling algorithm in a 1-core 1.7 GHz CPU is 20 seconds.


\subsection{Summarising the posterior PDF}
\label{posterior}

Having calculated the full posterior $p(M | O, B)$, we can integrate out
all parameters we are not interested in and leave out the PDF as a
function of only the oxygen and nitrogen abundances. The joint posterior
PDF for N/O and O/H (joint PDF for short) is calculated as
\begin{equation}
p( \{ z, n \} | O, B) = \sum_i p( \{ z_i = z, n_i = n, u_i, t_i, g_i \} | O, B ) \, \Delta u_i,
\end{equation}
where $z$, $n$, $u$, $t$, $g$ are the model input parameters log O/H,
log N/O, $\log U$, age and geometry, respectively, and the subscript $i$
tags each model in the grid. The sum is made over all models with the
same values of log O/H and log N/O, and $\Delta u_i$ takes into account
the variable octree cell size in $\log U$.  The expectation values for a
model input parameter $m_i$ or a computed emission line $c_j$ are given
respectively by
\begin{align}
E( \{ m_i \} | O, B) =
    {\cal N} 
     \sum_i
    \{ m_i \} \;
    p( \{ z_i = z, n_i = n, u_i, t_i, g_i \} | O, B ) \, \Delta u_i, 
\\
E( \{ c_j \} | O, B) =
    {\cal N} 
     \sum_i
    \{ c_j \} \;
    p( \{ z_i = z, n_i = n, u_i, t_i, g_i \} | O, B ) \, \Delta u_i.
\end{align}

The \bond code computes different summaries for the joint PDF: the
maximum a posteriori (MAP, i.e.\ the point of highest probability of the
joint PDF), the central point of the credible regions (i.e., the regions
on the N/O vs O/H plane of highest probability) that encompass 5, 50, 68
and 95 per cent of the joint PDF, plus its covariance ellipses (scaled
so that its area is the same as that of the credible region).

We also calculate the marginalised posterior PDF for several parameters
and emission lines. The marginalised posterior PDFs are summarised by
their average, median, and mode (i.e.\ the peak), plus their dispersion,
and the extremes of their 50, 68 and 95 percent equal-tailed (i.e.\
calculated from the percentiles) and highest density intervals.

For the sake of clarity, in what follows we show our results in three
descriptions only: the joint PDF, the maximum a posteriori (MAP) plus
the 68 per cent credibility ellipse, and the marginalised median plus
the 68 equal-tailed interval. Section~\ref{pdf-summaries} discusses the
differences in the summaries of the posterior PDF.


\subsection{A worked example}
\label{example}

\begin{figure*}
\centering
\includegraphics[width=\textwidth, trim=5 120 5 10]{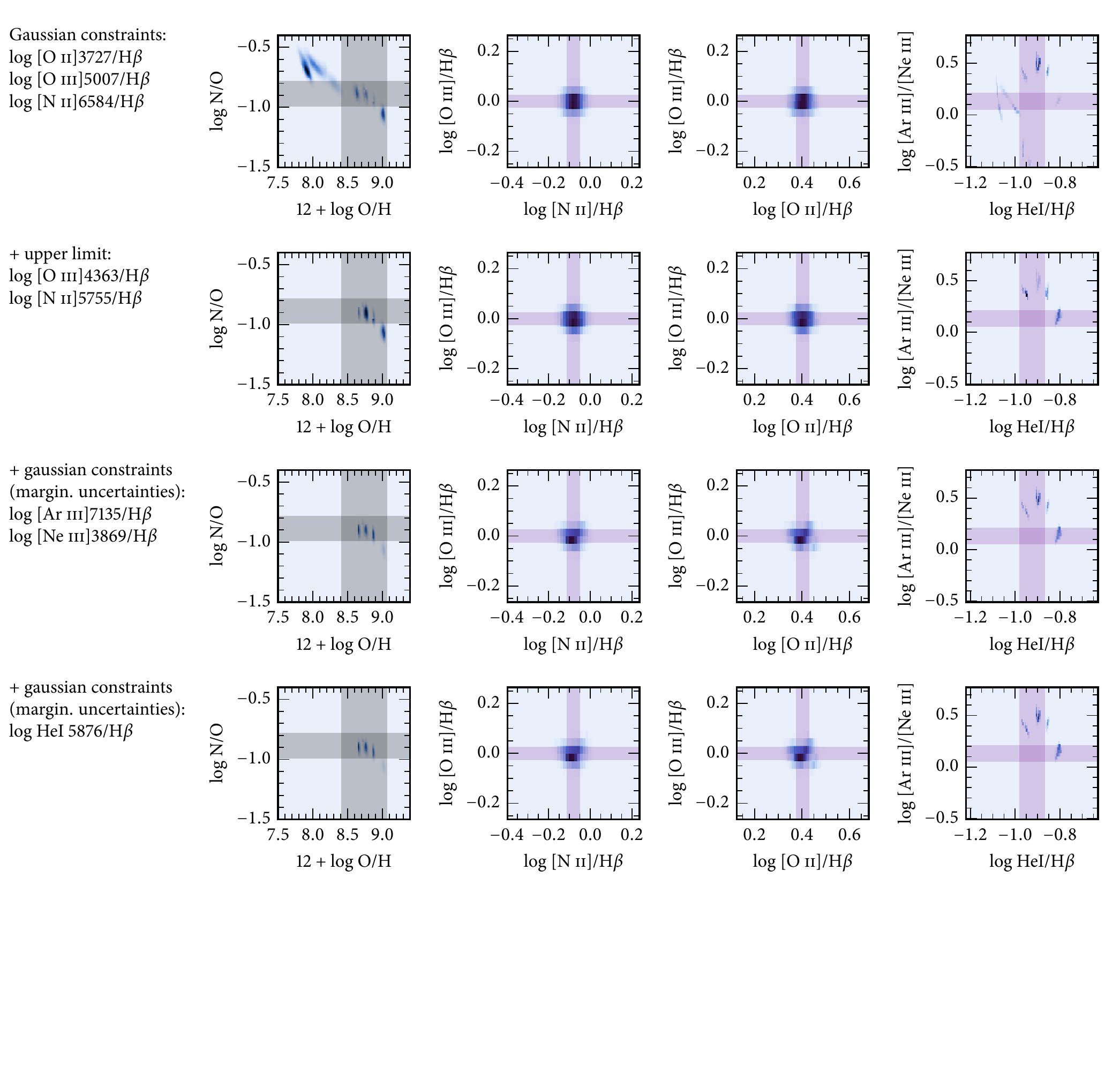}
\caption{Example of the \bond solution for an \hii region in NGC 1232
  (slit number 5 from \citealp{Bresolin.etal.2005a}).  The panels from
  left to right show the joint posterior PDFs in blue for N/O vs O/H,
  $\oiii/\Hb$ vs $\nii/\Hb$, $\oiii/\Hb$ vs $\oii/\Hb$, and
  $\ariii/\neiii$ vs $\hei/\Hb$.  The grey bands delimit $\pm 1 \sigma$
  of the temperature-based values for O/H and N/O, and the purple bands
  $\pm 1 \sigma$ of the observed line ratios.  Each row shows the effect
  of {\it cumulatively} adding another set of observational
  constraints. The top row shows the effect of a $\chi^2$ likelihood for
  $\oii/\Hb$, $\oiii/\Hb$ and $\nii/\Hb$. The second row applies the
  upper limit for $\Oiiit/\Hb$ and/or $\Niit/\Hb$, which selects the
  solutions on the high-metallicity branch. The third row imposes the
  constraint on $\ariii/\Hb$ and $\neiii/\Hb$, that further pins down
  the metallicity solution. Finally, the last row shows the effect of
  adding \hei/Hb, which selects all possible ionization sources. Note
  that the N/O vs O/H PDF is multipeaked, which means that there is a
  family of acceptable solutions in our grid (affecting mainly
  O/H). Those islands of solutions are a consequence of the discreteness
  of the starburst ages and nebular geometries in our grid.  }
\label{fig:pdfs}
\end{figure*}

\begin{figure}
\centering
\includegraphics[width=1.05\columnwidth, trim=20 38 25 20]{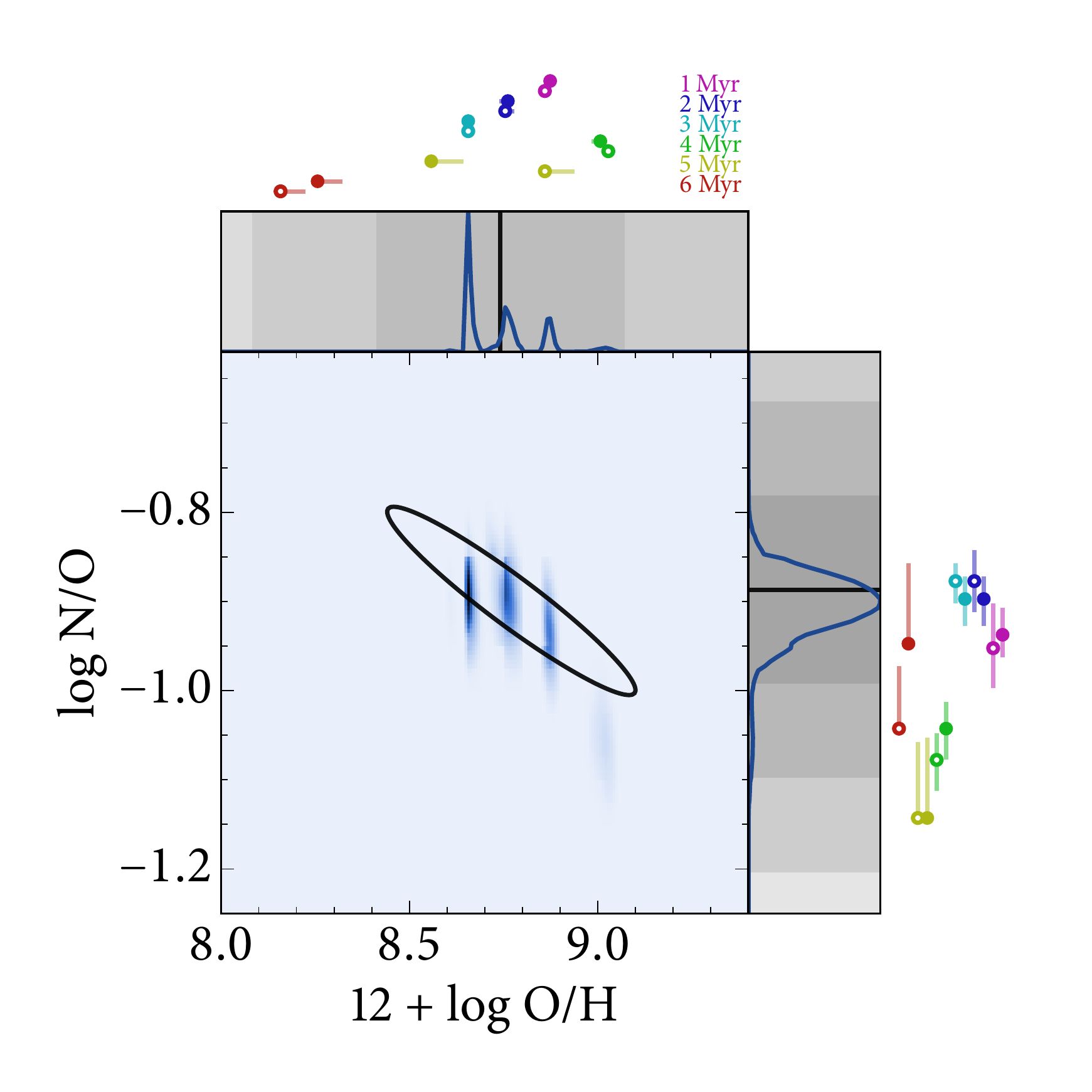}
\caption{A zoom in on the final N/O vs O/H panel for the same object in
  Fig.~\ref{fig:pdfs}, plus the marginalised PDFs for O/H and N/O.  This
  plot highlights how the discreteness of starburst ages in our grid of
  models leads to multipeaked solutions. We would expect a smoother PDF
  if the ages were more finely sampled.  Above the marginalised PDFs we
  mark the median (points) and the interval between the 16 to 84
  percentiles (lines) of the marginalised PDF for each age and geometry
  combination. Starburst ages are ordered from 1 to 6 Myr from top to
  bottom and colour-coded with a rainbow palette, while spherical shells
  and filled spheres are represented by open and filled circles,
  respectively.  The four islands of solutions on the joint PDF become a
  broad PDF in N/O and a multipeaked PDF in O/H.  To compare to the
  temperature-based method, we plot its covariance ellipse from the
  Monte Carlo realisations on the joint PDF, and mark the nominal
  temperature-based solution and $\pm 1,$ 2 and 3-$\sigma$ as grey bands
  on the marginalised PDFs. The nominal temperature-based solution is
  marked as a black line on the marginalised PDF panels.  }
\label{fig:marg_pdfs}
\end{figure}

Fig.~\ref{fig:pdfs} shows an example of the influence of each set of
constraints on the PDFs for an \hii region in NGC 1232.  Each row of
plots shows the effect of cumulatively adding a set of constraints to
the joint PDF of N/O and O/H and to the PDF of the emission line ratios
included in the likelihood calculation (\nii/\Hb, \oii/\Hb, \oiii/\Hb,
\ariii/\Hb, \neiii/\Hb and \hei/\Hb). The joint PDF is represented by a
blue-scale map.  The grey bands mark $\pm 1 \sigma$ of the
temperature-based values for O/H and N/O, and the purple bands delimit
$\pm 1 \sigma$ of the observed line ratios.

The top row shows the effect of using $\nii/\Hb$, $\oii/\Hb$,
$\oiii/\Hb$. A range of solutions pops up in the N/O vs O/H plane,
spread out in the high and low metallicity regimes.  The second row adds
the information on the upper limit detection of \Oiiit and \Niit. This
selects the high-metallicity solutions. The third row shows how
\ariii/\neiii weighs the four different `islands' of solutions.  Those
islands correspond to different starburst ages and geometries and are a
consequence of the sparsity of our grid in those parameters.  The last
row shows which solutions are favoured by \hei/\Hb, which will be those
with the right hardness for the ionizing source. For the \ariii/\Hb,
\neiii/\Hb and \hei/\Hb ratios, we add an extra noise source and then
integrate it out, to account for a dispersion in the Ar/Ne and He/H
abundance ratios in real nebulae with respect to the values adopted in
the model grid.

This example shows that there are three solutions of about the same
probability in the N/O vs O/H plane.  Fig.~\ref{fig:marg_pdfs} is a zoom
of the final N/O vs O/H plane, showing the joint and the marginalised
PDFs for N/O and O/H. The dots and lines above the marginalised PDFs are
the median and the interval between the 16 to 84 percentiles of the
marginalised PDF for each age and geometry combination. Those solutions
are displaced from the plot axes for clarity. Spherical shells and
filled spheres appear as open and filled circles, respectively, and ages
are colour-coded in the rainbow order with 1 Myr ages farther and 6 Myr
closer to the axes.

To compare to the Monte Carlo temperature-based realisations, we draw
its covariance ellipse on the joint PDF panel, and its $\pm 1$, 2 and
3-$\sigma$ as grey bands on the marginalised PDFs panels.  While the
method has not completely eliminated the multimodal nature of the
solutions, three out of the four islands of probability on the joint PDF
plane are compatible with the temperature-based solution (i.e. inside
its covariance ellipse).  The multimodal nature of the solutions draws
attention to the discreteness of our age grid, as marked by the coloured
dots and lines at the edges of the figure.  The important message from
this plot is that assuming different ionizing fields one finds different
values for O/H, and, while \hei/\Hb helps pinpointing the right ionizing
field, we may still end up with a range of acceptable solutions.  This
should improve when we have better constraints for the stellar radiation
field.


\section{Results}
\label{results}

\subsection
[The BOND N/O vs O/H diagram]
{The {\sevensize BOND} N/O vs O/H diagram}

\begin{table*}
  \caption{A sample of the posterior PDF summaries for sample B available for download
    at \url{http://bond.ufsc.br}. For log O/H and log N/O we report the
    maximum a posteriori (jmod); the centre, dispersion, covariance term
    and scaling to construct the 68\% credibility ellipse (jc68~cen, sig,
    cov, scale); and the marginalised median (mmed) and 68\%
    equal-tailed interval (mp68~low, mp68~upp).
    See Section~\ref{posterior} for details on how those terms
    have been defined.}
  \label{tab:bond}
\begin{tabular}{cccccccccccc}
\hline
id & name & log O/H & log N/O & log O/H & log O/H & \dots & log O/H & log N/O & log O/H & log O/H & \dots \\
 &  & jmod & jmod & jc68 cen & jc68 sig &  & mmed & mmed & mp68 low & mp68 upp &  \\
\hline
002 & NGC 1232 03 & $-3.7168$ & $-1.0797$ & $-3.6198$ & $0.0800$ & \dots & $-3.6359$ & $-1.0949$ & $-3.7195$ & $-3.5279$ & \dots \\
003 & NGC 1232 04 & $-2.8035$ & $-0.9974$ & $-2.8047$ & $0.0095$ & \dots & $-2.8041$ & $-0.9897$ & $-2.8180$ & $-2.7937$ & \dots \\
004 & NGC 1232 05 & $-3.3440$ & $-0.8872$ & $-3.2526$ & $0.0807$ & \dots & $-3.2655$ & $-0.9034$ & $-3.3454$ & $-3.1340$ & \dots \\
009 & NGC 1232 10 & $-2.9860$ & $-0.8719$ & $-2.9871$ & $0.0663$ & \dots & $-3.0012$ & $-0.8691$ & $-3.0258$ & $-2.9600$ & \dots \\
\dots & \dots & \dots & \dots & \dots & \dots & \dots & \dots & \dots & \dots & \dots & \dots \\
704 & F1629+205 & $-3.7860$ & $-1.3184$ & $-3.7247$ & $0.0693$ & \dots & $-3.7279$ & $-1.3333$ & $-3.7926$ & $-3.6240$ & \dots \\
706 & Mrk259 & $-3.8633$ & $-1.2782$ & $-3.7037$ & $0.1904$ & \dots & $-3.7960$ & $-1.2626$ & $-3.8678$ & $-3.4639$ & \dots \\
707 & SBS1428 & $-3.7087$ & $-1.1127$ & $-3.7678$ & $0.1419$ & \dots & $-3.7127$ & $-1.1080$ & $-3.9198$ & $-3.5843$ & \dots \\
708 & S1657+575 & $-3.8281$ & $-0.8720$ & $-3.8408$ & $0.1062$ & \dots & $-3.8013$ & $-0.8999$ & $-3.9260$ & $-3.6710$ & \dots \\
\hline
\end{tabular}
\end{table*}

We apply our \bond method to sample B, which contains 156 objects.
Table~\ref{tab:bond} shows a sample of the summaries for the posterior
PDF available to download from \url{http://bond.ufsc.br}.

\begin{figure*}
\centering
\includegraphics[width=0.85\textwidth, trim=0 10 0 10, clip]{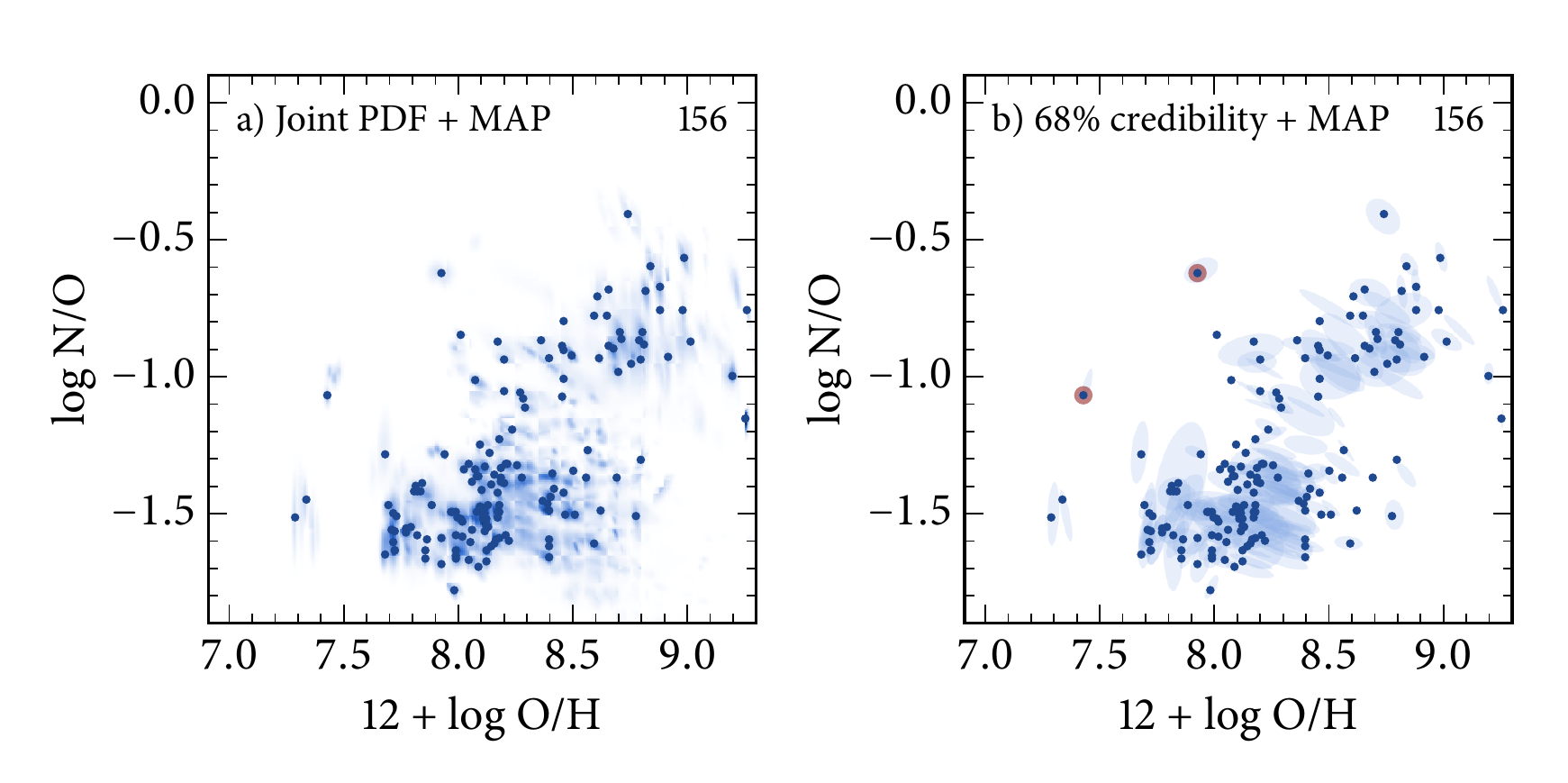}
\caption{Blue points on both panels are the maximum a posteriori (MAP)
  for each object in sample B. The blue scale on each map is a different
  description of our results.  (a) Superposition of the N/O vs O/H joint
  posterior PDFs for all objects. The blue gradient is linear and scaled
  such than the joint PDF for each object integrates to unity over this
  diagram.  (b) The ellipses are the covariances of the 68\% credibility
  region, and are scaled to cover the same area as the credibility
  region.  Strong multimodal solutions appear as MAP points far-flung
  from their elliptical companions.  Large red dots mark the two
  outliers on the N/O vs O/H plane (blue compact dwarves SBS0335-052E
  and 0837+4717).}
\label{fig:NOOH-BOND}
\end{figure*}

Fig.~\ref{fig:NOOH-BOND} shows the N/O vs O/H diagram obtained with
\bond for those objects.  The blue points in both panels are the maximum
a posteriori (MAP) values for each object. Panel a shows the
superposition of the joint PDFs for all objects, while panel b shows
the 68\% confidence ellipses. Note that some points fall a long way from
their ellipses; this is an evidence of multimodal solutions.

We find that our N/O vs O/H diagram is much more dispersed than the one
obtained by the ON method of \citet{Pilyugin.Vilchez.Thuan.2010a},
indicating that in nature this diagram is not a tight sequence. Points
are rather spread out like when using temperature-based
methods. Naturally part of the spread may be due to imperfections in our
models. For instance, we use a simplistic prescription for the nebular
density profile, which may not be realistic enough.

Because we do not impose any a priori solution for the N/O vs O/H
behaviour, unlike the tight correlation from the grid by
\citet{Blanc.etal.2015a} or the model selection by
\citet{PerezMontero.2014a}, we find outliers in the N/O vs O/H plane. In
Fig.~\ref{fig:NOOH-BOND} there are at least two objects with low O/H and
high N/O in both panels, marked as large red points in panel b
(SBS0335-052E and 0837+4717, id numbers 631 and 700 in
Table~\ref{tab:bond}). Those objects are likely interesting from a
chemical evolution perspective.

\subsection{The O/H ratio if one is not interested in N/O}
\label{pdf-summaries}

Given that we do not get rid of multimodal solutions, it is to be
expected that different ways to summarise the posterior (see
Section~\ref{posterior}) result in different solutions for O/H and N/O.
In theory, if one is not interested in N/O, one would be better advised
to use the mode, mean or median of the O/H PDF marginalised over N/O,
and those are not expected to be exactly the same as the descriptions
for the joint PDF.

\begin{figure}
\centering
\includegraphics[width=0.8\columnwidth, trim=0 10 0 10, clip]{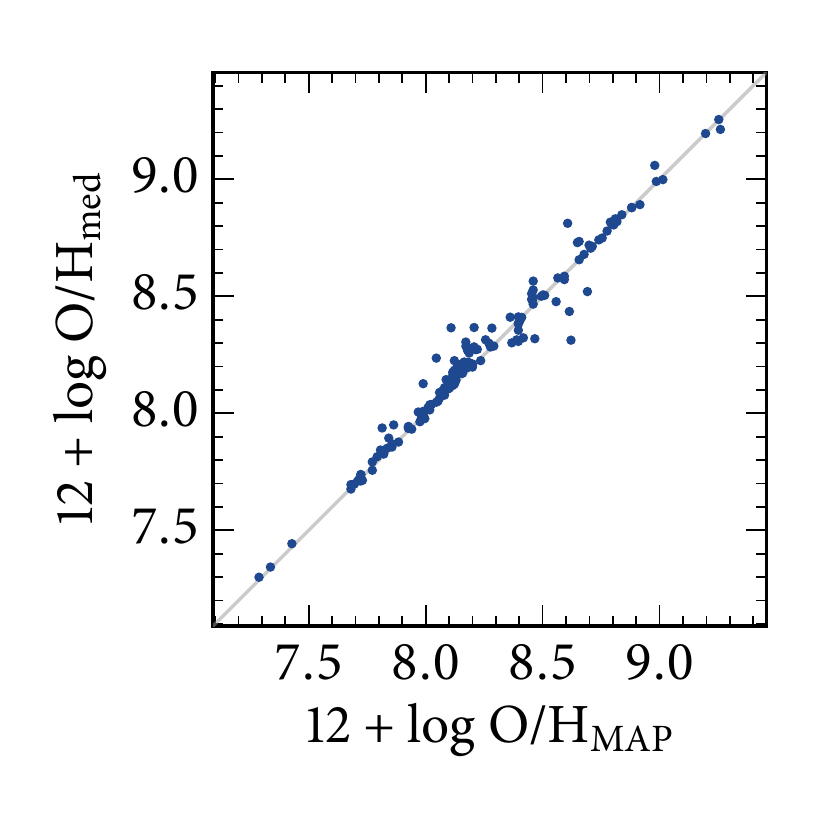}
\caption{Comparison between the maximum a posteriori O/H and the
  marginalised median O/H. The identity line is drawn in grey.  }
\label{fig:OH-jpdf-mpdf}
\end{figure}

Fig.~\ref{fig:OH-jpdf-mpdf} compares the O/H from maximum a posteriori
(MAP) value of the joint N/O vs O/H PDF to the median of the
marginalised O/H PDF. For most objects those two solutions agree to
within 0.1 dex: There is no bias between those two nominal solutions
(the average difference is $< 0.02$ dex) and the dispersion in small
($0.06$ dex). Other descriptions of the joint and marginalised PDFs may
have a better or worse agreement, but the important point is to always
test a few of those descriptions plus their credibility regions.


\subsection{Comparison to the direct method}

\begin{figure*}
\centering
\includegraphics[width=0.7\textwidth, trim=0 10 0 10, clip]{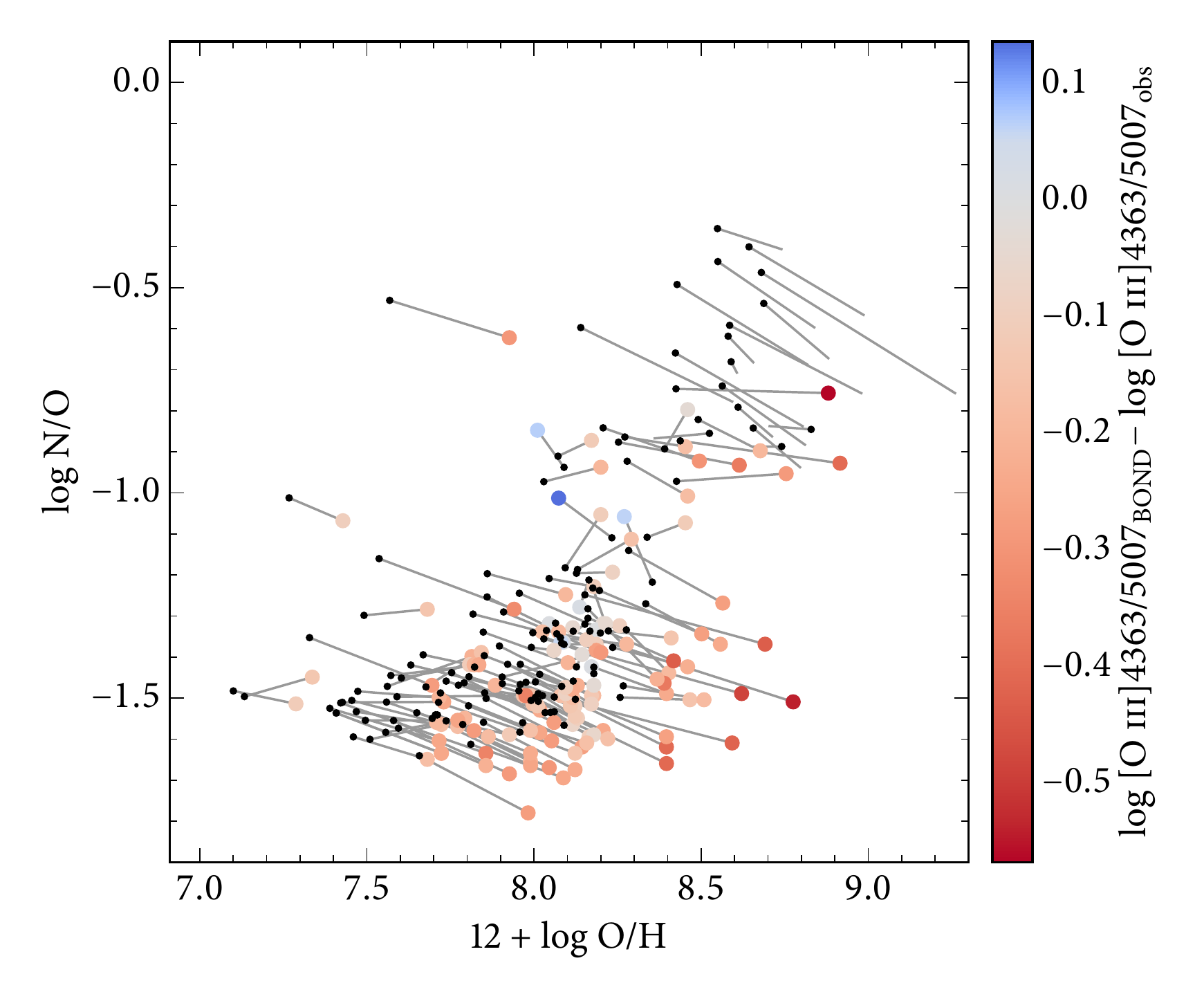}
\caption{Comparison of N/O vs O/H as obtained with the temperature-based
  method (small black circles) and with the \bond method
  (large coloured circles). Solutions for the same object are
  joined by a line.  The large circles are coloured with
  respect to the difference between \Oiiit/\Oiii for the fitting models
  and the measured \Oiiit/\Oiii. Lines lacking a coloured
  counterpart indicate there is no measured \Oiiit (i.e.\ the
  direct method used \Niit/\Nii to measure the electron
  temperature). Note how points tend to shift to higher O/H with \bond,
  and that the temperature in the models tend to be smaller than the one
  inferred from the auroral lines. A colour version of this
    figure is available in the electronic edition.}
\label{fig:Te-bond}
\end{figure*}

Fig.~\ref{fig:Te-bond} shows a comparison of \bond to temperature-based
results for the objects in sample B that have a direct temperature
measure. The small black circles are the temperature-based results and
the large coloured dots are the \bond results. The latter are
colour-coded according to the difference between log \Oiiit/\Oiii in the
models and the observed value (blue for extreme positive values to red
for extreme negative values).  The \bond and temperature-based results
for the same \hii region are joined by a solid line.

We see that the \bond results generally migrate towards higher values of
O/H and lower values of N/O.  The colour-coding shows that, for the
majority of the points, the \bond models have lower temperatures than
the observations. This explains why the \bond oxygen abundances are
higher than the temperature-based ones (typically by 0.2--0.4
dex). Concomitantly, the \bond N/O ratios are smaller since the
emissivity of the \nii line is less dependent on the temperature than
that of the \oii line, which has a higher excitation threshold.

This problem is not unique either to our grid of models or to our
code. \citet{PerezMontero.2014a} only obtains O/H values that are in
agreement with the direct method when \Oiiit\ is fitted (see his
fig.~2). His method gives a huge weight to \Oiiit/H$\beta$ (see his
eq.~19); in other words, it becomes essentially a temperature-based
method.  \citet{Blanc.etal.2015a} compared the results from their code
\izi using several photoionization models
\citep{Kewley.etal.2001a,Levesque.Kewley.Larson.2010a,Dopita.etal.2013a},
and found offsets of $-0.07$ to $0.32$ dex with respect to recombination
lines, which translate into 0.17--0.56 dex offsets with respect to the
temperature-based method.

It is a well-known problem that collisionally excited lines, when using
temperature-based methods, lead to lower oxygen abundances than
recombination lines \citep[see e.g.][]{GarciaRojas.Esteban.2007a},
typically by 0.2--0.3 dex in \hii regions. One explanation for this
abundance discrepancy problem could be that, in real nebulae, strong
temperature fluctuations (\citealp{Peimbert.1967a}; more important than
the temperature gradients arising in classical photoionization models)
or a $\kappa$ distribution of the free electron velocities
\citep{Nicholls.Dopita.Sutherland.2012a} boost the \Oiiit line, a fact
which is not taken into account in the classical temperature-based
method nor in the grid of photoionization models we use.

The simplicity of the direct method can be a double-edged sword:
although powerful and straightforward to apply, it might be holding too
simplistic assumptions as to the production of \Oiiit.

\section{Summary, discussion and future directions}
\label{summary}

\bond determines nitrogen and oxygen gas-phase abundances by using
strong and semi-strong lines and comparing them to a grid of
photoionization models in a Bayesian framework. The code is written in
python and its source is publicly available at
\url{http://bond.ufsc.br}. The grid of models presented here is included
in the 3MdB database \citep*[][see
\url{https://sites.google.com/site/mexicanmillionmodels/}]{Morisset.DelgadoInglada.FloresFajardo.2015a}
under the reference `BOND'. The Bayesian posterior probability
calculated by \bond stands on two pillars: our grid of models and our
choice of observational constraints (from which we calculate our
likelihoods). We discuss each of these in turn.

The ideal grid of models should be all-encompassing and able to describe
any emission-line object found in nature. Creating such a grid would be
a daunting and nearly impossible task; therefore we have crafted a set
of models that covers enough physical parameters of \hii regions not to
be plagued by the usual preconceptions that go into making these grids.
Our models span a wide range in N/O and O/H, without imposing any
relation between N/O and O/H. The only model grid that has so far taken
this approach is the one by \citet{PerezMontero.2014a}. Unlike his
method, we leave the starburst age and the nebular density structure as
free parameters.  Finally, a crucial step forward in our approach is
taking into account the importance of the hardness of the ionization
field. All model grids in the literature consider only a single type of
ionizing sources. If the ionization field of the \hii regions differs
from those in the models, the O/H obtained will be strongly biased (see
the discussion in Appendix~\ref{tests}).  The hardness of the ionization
source may vary due to a few reasons. In local galaxies, the main
effects will be the ageing of the stellar populations in \hii regions,
and the stochastic sampling of the stellar initial mass function
\citep[e.g.][and references therein]{Cervino.etal.2013a}.  Our model
grid attacks this by using simple stellar populations (SSPs) of
different ages for the ionizing sources to account for variations in the
ionizing field.

Constructing the grid with great care is not enough. Given the strengths
and limitations of our set of models, we need to critically assess which
theoretical predictions should be trusted and which observational
constraints should go into our fitting procedure. We have set out to
infer O/H and N/O, but also to constrain the nuisance parameters $U$,
the correct O/H bimodality branch, and the hardness of the ionizing
field. The strong lines \oiii/\Hb, \oii/\Hb and \nii/\Hb constrain O/H,
N/O and $U$. To pin down the correct O/H branch, we use an upper limit
in \Oiiit or \Niit when at hand, and the ratio of the semi-strong lines
\ariii/\neiii, which depends mostly on the electron temperature
(\textit{modulo} the ionizing structure of the nebula, already
constrained by \oiii/\oii). Indeed \ariii and \neiii have different
excitation thresholds while Ar/Ne in the gas phase is expected to be
constant, since both argon and neon are primary elements and
inert. Lastly, another semi-strong line comes to rescue: \hei/\Hb helps
constrain the mean effective temperature of the ionizing radiation
field.

Unlike several authors \citep{Dopita.etal.2013a, PerezMontero.2014a,
  Blanc.etal.2015a} we do not use the \Sii line in our procedure, first
because this line is emitted in the outskirts of \hii regions so that
its intensity in relation with \oii or \nii is dependent on the detailed
density structure of the nebulae, second because there might be an
intrinsic scatter in the S/O ratio due to stellar nucleosynthesis and/or
to depletion effects not yet fully documented.

For a set of giant \hii regions and blue compact dwarf galaxies, we have
calculated their gas-phase N/O and O/H abundances and compared them to
the ones obtained by the temperature-based method and by the
\citet{Pilyugin.Vilchez.Thuan.2010a} ON method. We find that the N/O vs
O/H relation obtained by \bond is as scattered as the one obtained by
the temperature-based method, and that the very tight relation obtained
with the Pilyugin method is a consequence of that method itself.

We also note that, when using the \bond method on objects which have
direct temperature measurements, we systematically obtain lower values
of \Oiiit/\Oiii than observed and higher values of the oxygen abundance
than with the temperature-based method. This discrepancy has been seen
in many other strong-line methods calibrated on photoionization models
and might point to too soft an ionizing spectral energy distribution
(SED) in the models. This is in line with the fact that
\citet{Stasinska.etal.2015a}, using a subset of our models, found that
the SEDs are not hard enough to produce the observed \ariv/\ariii line
ratio. It might also indicate that the density distributions of our
models are too simplistic to represent real \hii\
regions. Alternatively, it might be a sign that important temperature
fluctuations of unknown origin or a $\kappa$ distribution of electron
velocities are present in real \hii regions and lead to an overestimate
of the temperature indicated by \Oiiit/\Oiii. Such an explanation has
been suggested by several authors and would at the same time help
resolve the famous abundance discrepancy problem \citep[see
e.g.][]{GarciaRojas.Esteban.2007a, Nicholls.Dopita.Sutherland.2012a,
  Dopita.etal.2013a}. Since we do not try reproduce the \Oiiit line,
nitrogen and oxygen abundances inferred by \bond might be more accurate
estimates than those of the temperature-based method. Nevertheless, the
accuracy of \bond abundances strongly relies on how well the
photoionization model grid represents real objects.

We have shown that \bond, when applying our extensive grid of
photoionization models to a well-chosen set of strong and semi-strong
lines, allows one to obtain O/H and N/O simultaneously, getting rid of
the spectre of bimodality without recourse to empirical oxygen and
nitrogen abundance correlations.  Our method is very easily extendable
and can accommodate many improvements in the future.  In spite of many
issues still to resolve in the determination of nebular abundances, we
hope that \bond does offer a quantum of solace.

\section*{Acknowledgments}  

We are thankful to Robert Kennicutt and the referee, Guillermo Blanc,
for their insightful comments, which led us to rephrase a number of
issues to improve clarity. We also thank Guillermo Blanc for the
kindness of running \izi for us in the refereeing process. GS and NVA
acknowledge the support from the CAPES CsF--PVE project
88881.068116/2014-01. NVA acknowledges the NEBULATOM school held in
Choroni, Venezuela, in 2013, and the support and hospitality for short
term visits of the LUTH, at Observatoire de Paris. GS and RCF
acknowledge the support from the CAPES-COFECUB Te 585/07 project. The
grid of models has been run on computers from the CONACyT/CB2010:153985,
UNAM-PAPIIT-IN107215 and UNAM Posgrado de Astrof\'{i}sica projects. This
research made use of Astropy, a community-developed core Python package
for Astronomy; matplotlib, a Python library for publication quality
graphics; SciPy; numpy; h5py; and hickle.

\bibliography{references}


\appendix


\section{A Fake sample for tests}
\label{fake}

\begin{figure*}
\centering
\includegraphics[width=0.7\textwidth, trim=0 10 0 10, clip]{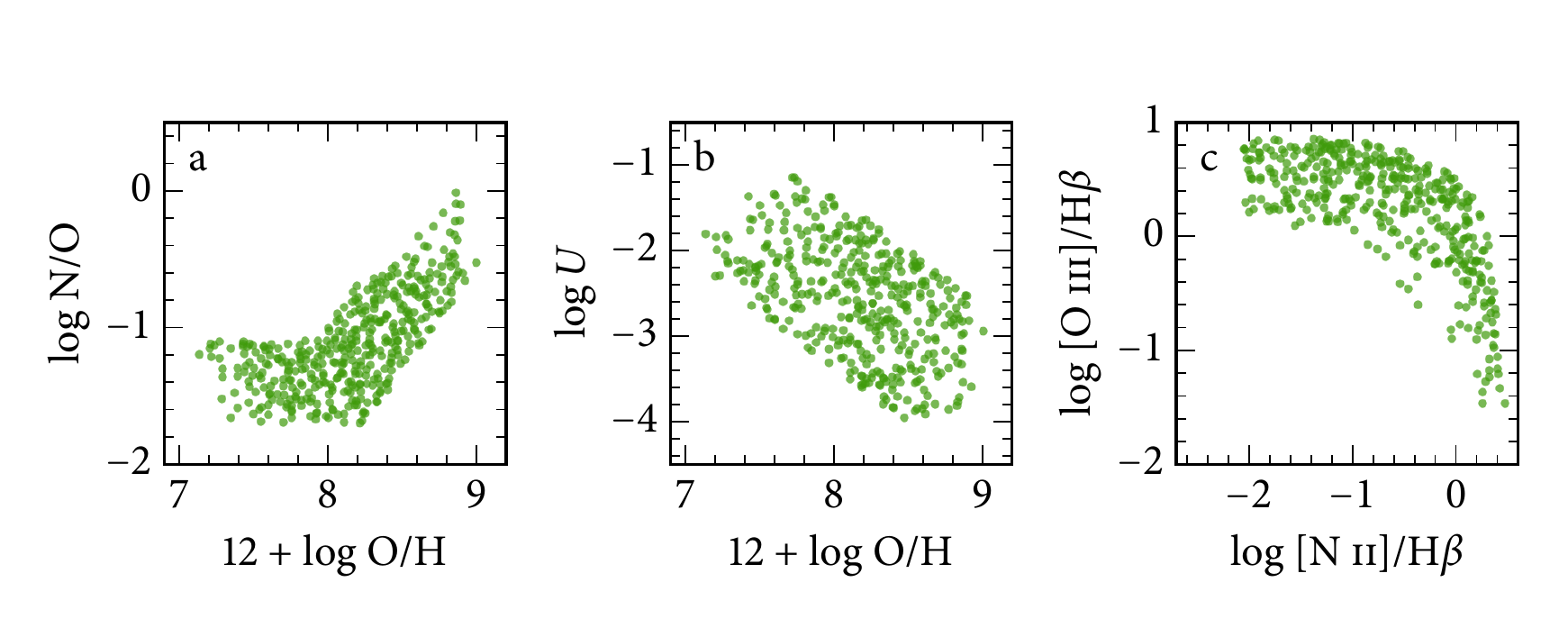}
\caption{ Fake sources chosen from the grid with age 2 Myr and filled
  sphere.  The grid is finer than our original grid, but not as fine as
  the interpolated one we use as our initial octree grid. The panels
  show our selection of grid models on the N/O vs O/H, $U$ vs O/H and
  \citetalias{Baldwin.Phillips.Terlevich.1981a} planes.}
\label{fig:fake}
\end{figure*}

\begin{figure}
\centering
\includegraphics[width=\columnwidth, trim=20 16 20 10, clip]{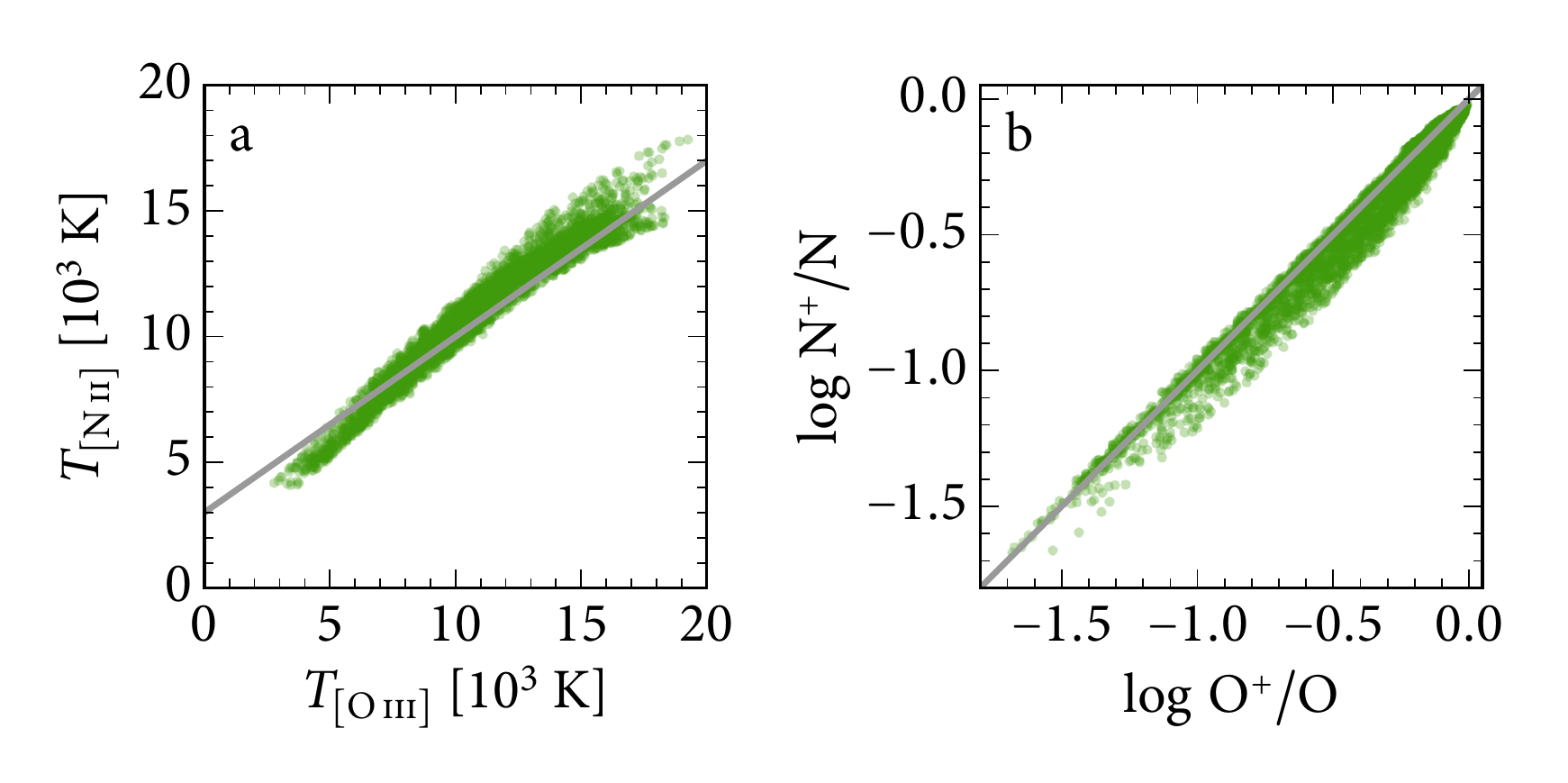}
\caption{ (a) Comparison between the temperatures in the high and low
  excitation zones for the fake sources. The grey line is the classical
  relation from \citet{Garnett.1992a} given by Eq.~\ref{eq:T}.  (b)
  Comparison between the ionic fractions of \Np and \Op. The one-to-one
  line is in grey.  }
\label{fig:fakeT}
\end{figure}

To assess some of the recipes used in temperature-based methods and to
test the \bond method on objects with known abundances, we construct a
`fake' sample by selecting from our grid of models a subsample that
roughly follows the expected properties of our observational sample.

To limit the size of our fake sample, we have created an interpolated
grid having a resolution of 0.1 dex in O/H and N/O, and 0.5 dex in $U$.
We then perturb each cell point in the grid with uniform noise (setting
its maximum amplitude to be the size of the cell) in those three input
parameters, so that the fake sources are not superimposed in our plots.
Our `fake' sources are then chosen to fall roughly in the same loci as
observed data, as Fig.~\ref{fig:fake} shows.  We select models in the
vicinity of the observed N/O and O/H relation as expressed by eq.~2 of
\citealp{Pilyugin.etal.2012b}, around the $U$ vs O/H relation found by
\cite{PerezMontero.2014a}, and below the \citet{Stasinska.etal.2006a}
line delimiting pure \hii regions in the
\citetalias{Baldwin.Phillips.Terlevich.1981a} diagram.  For a given age
and geometry we have around $\sim 350$--$400$ fake sources (except for
the 6 Myr scenarios, which fail to cover a large part of the
observational data and thus have $\sim 160$ fake sources).

Fig.~\ref{fig:fakeT}a shows the relation between the temperatures in the
high and in the low excitation zones and Fig.~\ref{fig:fakeT}b shows the
relation between the ionic fractions of \Np and \Op. The temperatures
and ionic fractions come directly from the photoionization models for
our fake source sample.  The continuous lines indicate the relations we
used in the temperature-based method. We see that they represent well
the trends shown by the models. We also see that the models show some
dispersion about these lines. For the temperature the dispersion is of
600 K, while for the logarithm of ionic abundances it is of 0.06 dex.

\section{Tests of the accuracy of the BOND method}
\label{tests}

Here we run a suite of tests fitting models with models, using the same
code and the same assumptions as for the sources in sample B. The aim of
this exercise is twofold. First, we show that our method works when the
input and the output are the same, which is the zeroth test of
reliability of any method. Second, we check how the different ionizing
source ages and density structures affect our results, since this is the
main novelty of our model grid.

The model grid considered is the octree sampled grid. For the tests in
this section we select subgrids of single ages and density structures to
highlight the effects of those parameters.  For all the tests presented
here, we assume that the uncertainties in the intensity ratios
\oiii/\Hb, \oii/\Hb, \nii/\Hb, \ariii/\Hb, \neiii/\Hb, and \hei/\Hb are
of 10\%.

\begin{figure}
\centering
\includegraphics[width=\columnwidth, trim=30 20 260 0, clip]{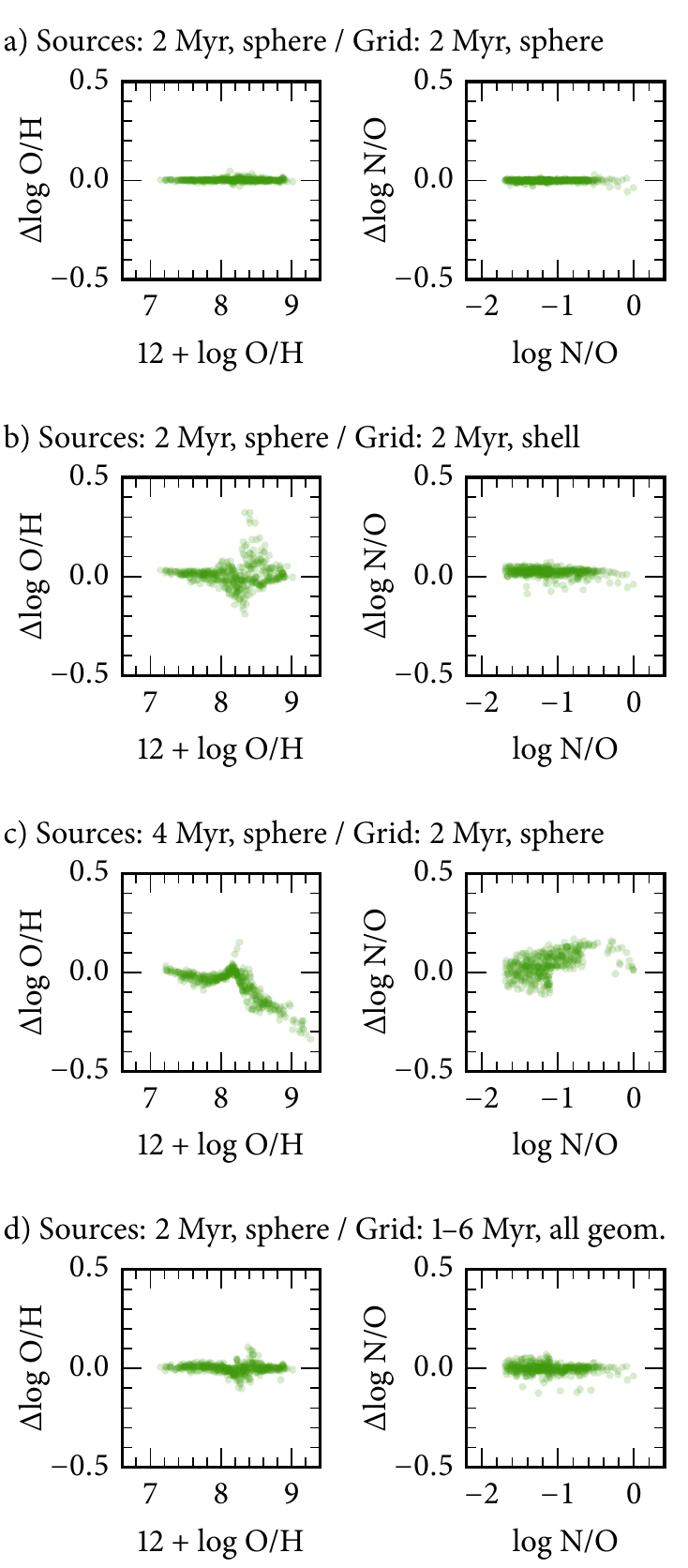}
\caption{Results from modelling the grid fake sources with \bond. The
  panels show the residual parameters (input minus output) for O/H and
  N/O.  The top row shows that we recover O/H and N/O quite well when we
  fit sources from the 2 Myr filled sphere grid with a grid of the same
  age and geometry. The second row swaps the filled sphere by a shell in
  the fitting grid.  The third row fits 4 Myr fake sources with a grid
  of the same geometry but 2 Myr ionizing sources. We see that O/H is
  highly biased for the high-O/H solutions.  The last row shows 2 Myr
  filled spheres fitted with our entire grid. The results for O/H are
  very good, with a dispersion of only 0.02 dex.}
\label{fig:gridgrid}
\end{figure}

The results of our tests are shown in Fig.~\ref{fig:gridgrid}, where the
different rows correspond to different choices of age and geometry.  The
first row shows 2-Myr starburst and spherical shell fake sources modelled
with a grid of the same age and geometry. For O/H and N/O, we show the
difference $\Delta$ between the output and the input as a function of
the input parameter. The results shown are for the maximum a posteriori
values. N/O and O/H are well recovered (the dispersion is 0.007 dex).


The second and third rows of Fig.~\ref{fig:gridgrid} show the effect of
using the wrong density structure and age, respectively. The second row
shows the 2 Myr filled sphere fake sources fitted with 2 Myr spherical
shell models.  The residuals for O/H and N/O are very dispersed (0.02
and 0.06 dex respectively) and slightly biased (0.006 and 0.02 dex). The
third row shows the effect of using the wrong hardness for the
ionization source. Here we have 4 Myr fake sources fitted with a 2 Myr
grid, both modelled as filled spheres. This is a very worrying scenario:
O/H is underestimated by $\sim 0.1$ dex (and up to 0.4 dex) for the
high-metallicity branch, while N/O is slightly overestimated (0.04 dex)
and rather dispersed (0.05 dex).

The last row shows how our 2-Myr filled sphere fake sources are modelled
using our entire grid, i.e. without any a priori knowledge of the
geometry and the age of the ionizing source.  The results are quite
encouraging, and the code seems to choose the right age and geometry
combination; or, at least in practice, the right O/H and N/O solution.
O/H and N/O are recovered to within better than 0.05 dex (0.02 dex of
dispersion).


\section{Comparison with other strong line methods}
\label{comparison}

\begin{figure}
\centering
\includegraphics[width=0.9\columnwidth, trim=10 30 30 70, clip]{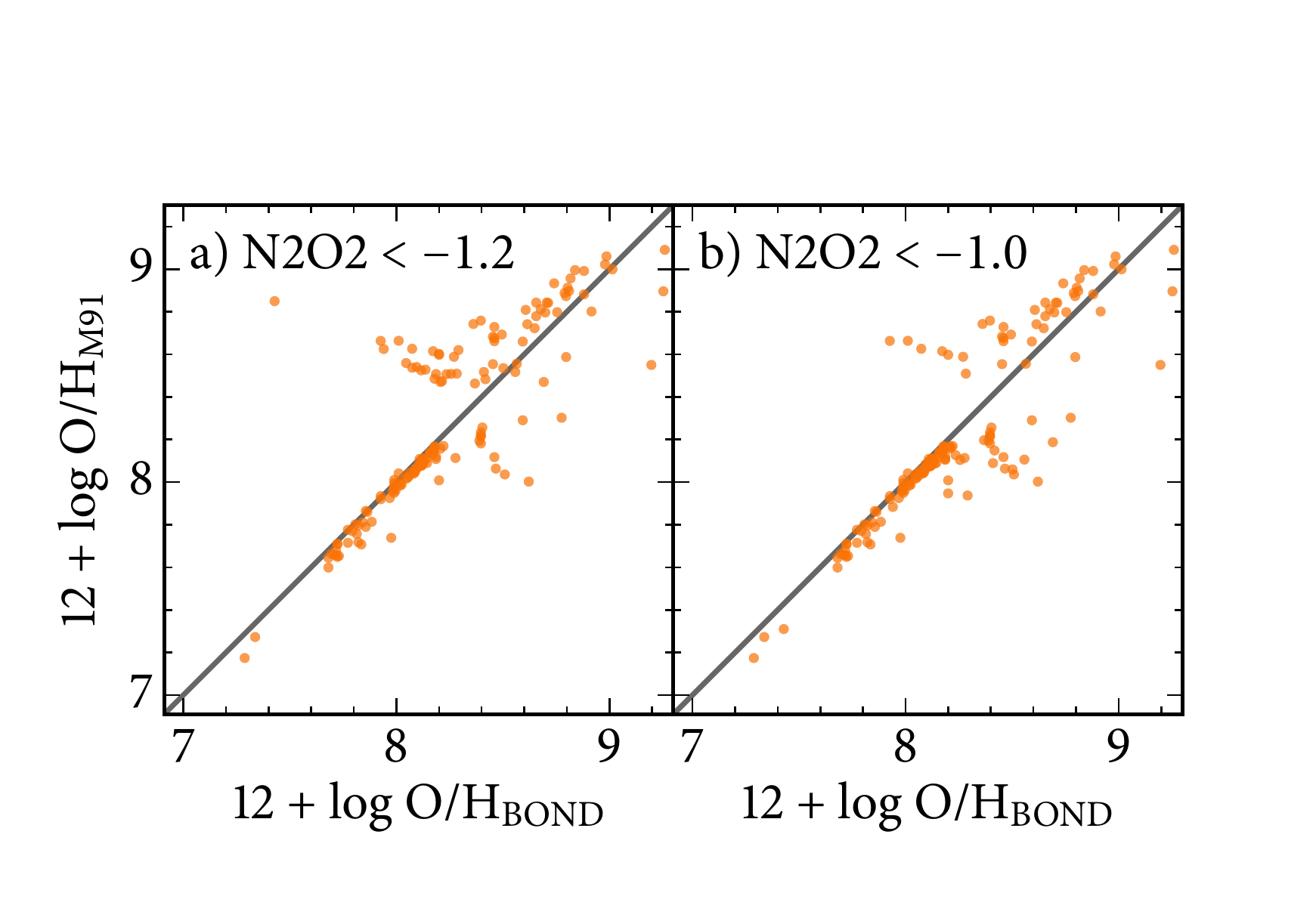}
\caption{Comparison between O/H from \bond and from \citet{McGaugh.1991a} for our
  sample B. The panel on the right shows O/H from McGaugh by choosing
  the O/H according to $\log \, \nii/\oiii = -1$ \citep{McGaugh.1994a}, and on
  the left according to  $\log \, \nii/\oii = -1.2$ \citep{Kewley.Ellison.2008a}.}
\label{fig:m91}
\end{figure}

Here we compare our \bond method to several other strong line
methods. Fig.~\ref{fig:m91} shows the comparison to the O/H measured by
\citet{McGaugh.1991a}, using eqs.\ A1 and A2 from
\citet{Kewley.Ellison.2008a}. The two panels show the effect of choosing
different criteria to separate the low and high-metallicity solutions:
On the right we use $\log \, \nii/\oii = -1$ \citep[as
in][]{McGaugh.1994a}, and on the left $-1.2$ \citep[as
in][]{Kewley.Ellison.2008a}.  The separation between the two branches is
fuzzy, and the effect of choosing a slightly different frontier is seen
when comparing one panel to the other.  Focusing on the comparison of
\citet{McGaugh.1991a} to \bond, we see a good agreement between the
results from \bond and those from the much simpler McGaugh recipe, but
there are important differences for a non negligible number of sources
at high metallicities. There are, as expected, huge differences around
$12 + \log \mathrm{O/H} = 8.5$, where the O23 ratio is insensitive to
metallicity while \bond is aided by using the \ariii/\neiii ratio, which
steadily increases with increasing metallicity.

\begin{figure}
\centering
\includegraphics[width=0.9\columnwidth, trim=10 30 30 70, clip]{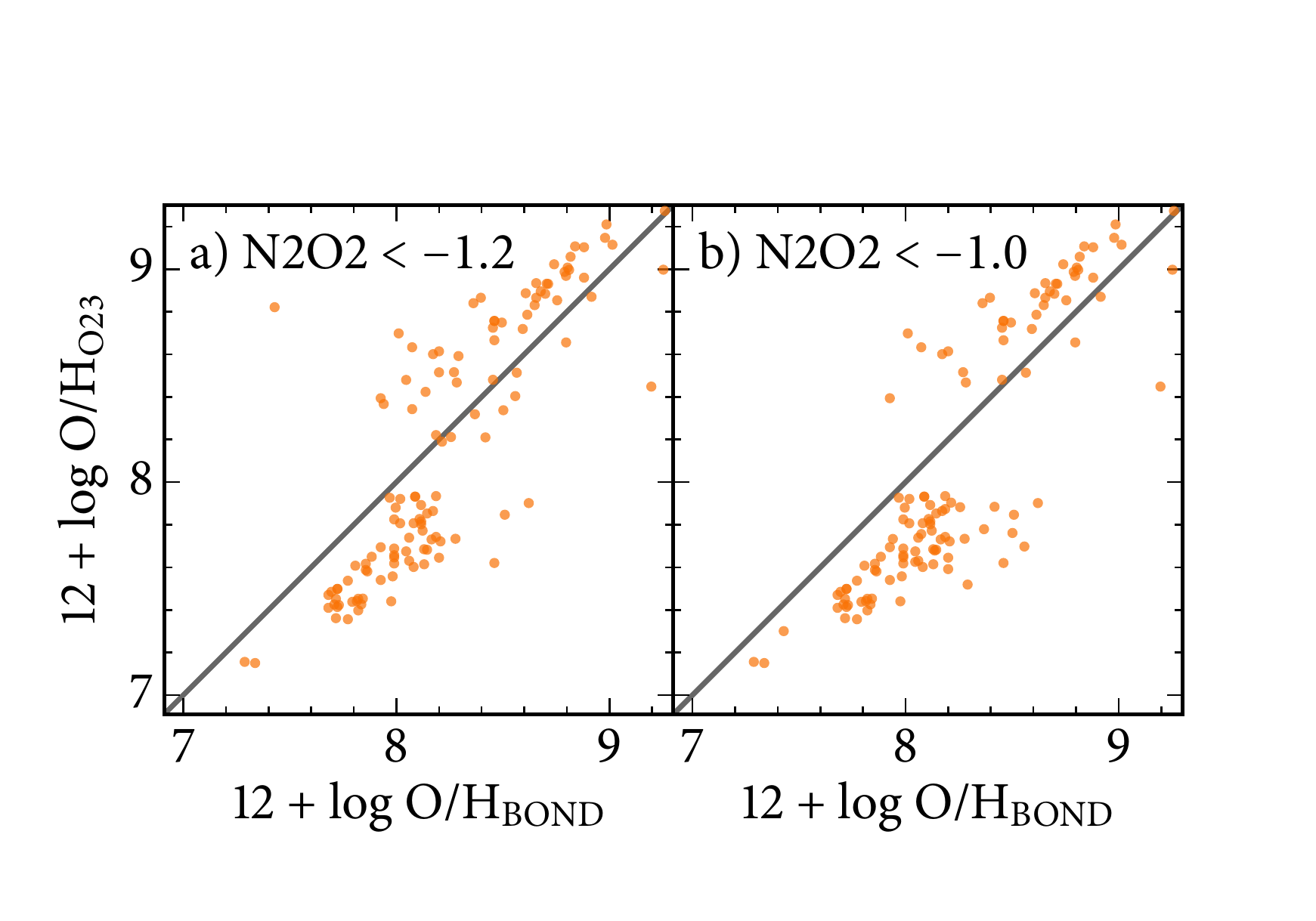}
\caption{As Fig.~\ref{fig:m91}, but for O/H derived with the
    O23 calibration from \citet{Maiolino.etal.2008a}.}
\label{fig:O23-M08}
\end{figure}

\citet{McGaugh.1991a} was the first to take into account the effect of
the ionization parameter when measuring abundances.
Fig.~\ref{fig:O23-M08} shows the comparison of \bond results with those
using a simple O23 calibration (we have used the one by
\citealp{Maiolino.etal.2008a} as an example). The systematics with this
simple O23 calibrator are of the order of 0.2--0.5 dex over the whole
O/H range.

\begin{figure}
\centering
\includegraphics[width=\columnwidth, trim=10 20 10 20]{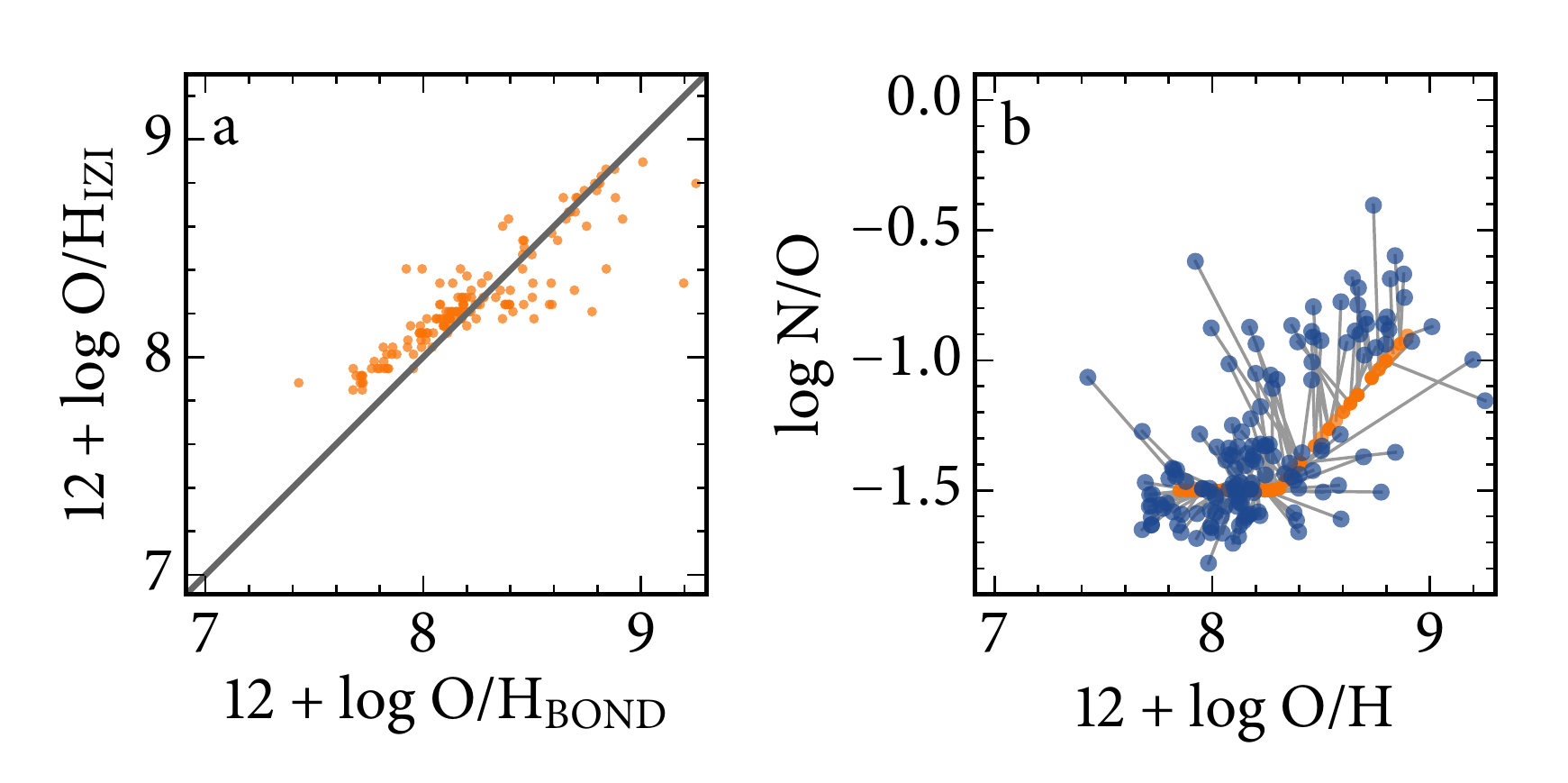}
\caption{How the assumption of a relation between N/O and O/H affects
  the derived O/H. Panel a compares the values of O/H obtained by \bond
  to those by \izi \citep{Blanc.etal.2015a} with their default grid,
  which assumes a relation between N/O and O/H. We show the results for
  151 objects in our sample B that have also been flagged as having
  reliable results by \izi. Panel b shows N/O as a function of O/H for
  \bond and \izi. The results using the \citet{Blanc.etal.2015a} code
  and grid are the small orange points, and those with \bond are large
  blue points. Results for the same object are linked by a grey line. }
\label{fig:blanc}
\end{figure}

Fig.~\ref{fig:blanc} compares the \bond results obtained with our full
grid of models to those from the \izi code by \citet{Blanc.etal.2015a}
with their default grid (\citealp{Levesque.Kewley.Larson.2010a}, 6 Myr
constant star formation).  \citet{Blanc.etal.2015a}, like
\citet{McGaugh.1991a}, assume a N/O vs O/H relation and consider a
unique family of ionizing stellar energy distributions. Panel a shows
the comparison between the values of O/H derived by \bond and \izi for
the 151 objects in sample B (see Sec.~\ref{sec:subsamples}) which have
the \izi quality flags \texttt{npeakZ} and \texttt{limZ} equal to one
(this removes only 5 objects from sample B).  For
$\log \mathrm{O/H} \lesssim 8.4$, \izi metallicities are systematically
larger than \bond by 0.1--0.2 dex. For $\log \mathrm{O/H} \gtrsim 8.4$,
the O/H from \bond can be 0.3--0.8 dex larger \izi for a few objects,
while for other objects the codes agree quite well (differing by
$\lesssim 0.02$ dex).  Panel b of Fig.~\ref{fig:blanc} shows the values
of N/O vs O/H derived by \bond linked by a straight line to the values
obtained by \citeauthor{Blanc.etal.2015a} for the same objects
(actually, \citeauthor{Blanc.etal.2015a} determines only O/H, since the
N/O values lie on the relation assumed by them). We see that some
objects are actually quite far from the tight N/O vs O/H relation
assumed, and that for those objects the O/H values derived by \bond
differ substantially from those derived by
\citeauthor{Blanc.etal.2015a}. This illustrates that, for a number of
objects (which are not the majority but are not known a priori) it is
necessary to simultaneously derive N/O and O/H to obtain a reliable
oxygen abundance.

\begin{figure*}
\centering
\includegraphics[width=0.9\textwidth, trim=0 30 0 10]{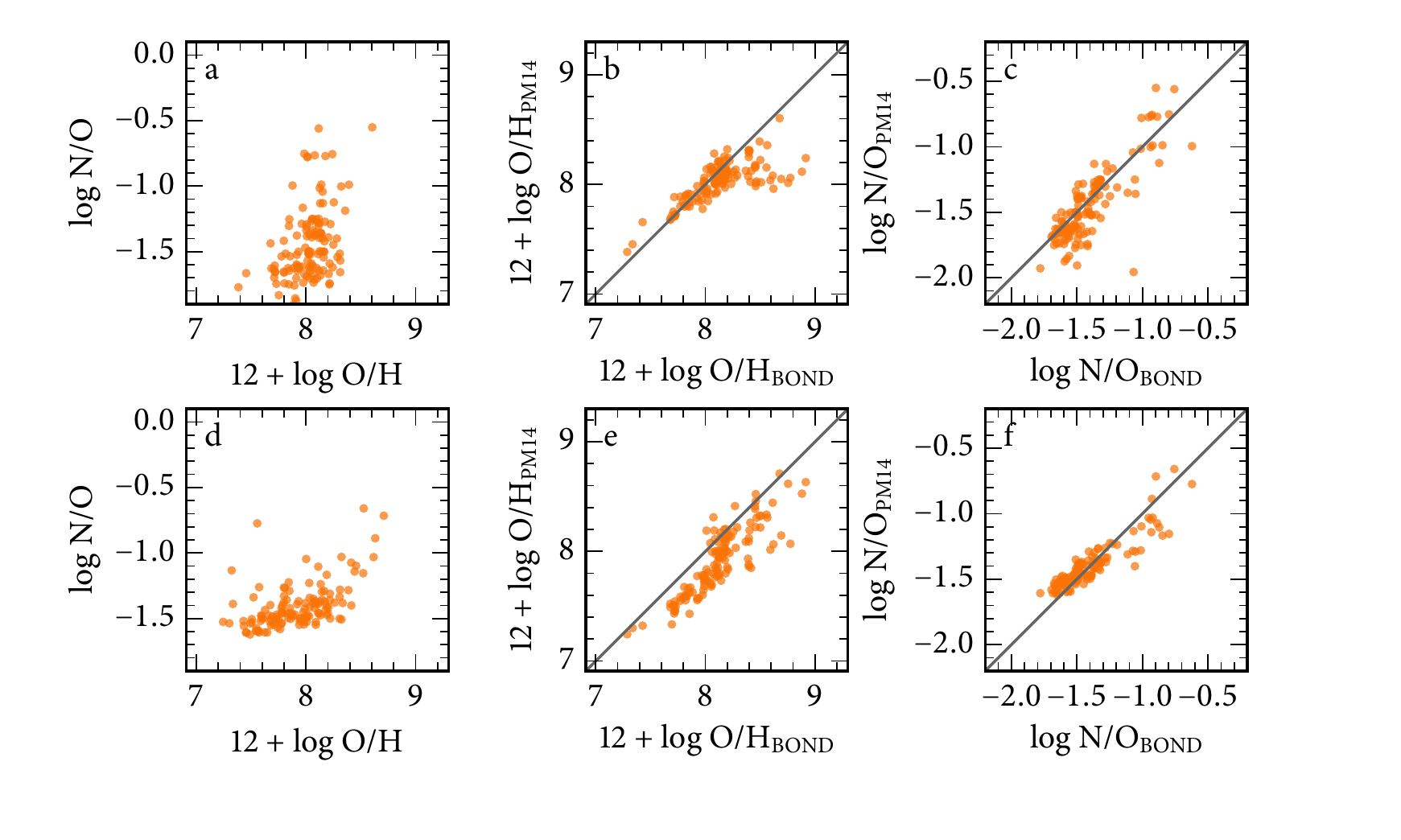}
\caption{Results for applying the {\scshape hii-chi-mistry} code by
  \citet{PerezMontero.2014a} to our sample B. Top and bottom
  rows respectively show results excluding or including \Oiiit
  from the {\scshape hii-chi-mistry} fits. From left to right, the N/O
  vs O/H diagram, and the comparison between the O/H and N/O
  values to \bond.}
\label{fig:PM14}
\end{figure*}

Fig.~\ref{fig:PM14} shows the comparison between \bond and the {\scshape
  hii-chi-mistry} code by \citet{PerezMontero.2014a} for our sample
B. We do not show objects that the \citet{PerezMontero.2014a} code flags
as bad, i.e.\ when his {\tt grid} output is either 2 or 3. From the
original 156 objects in sample B, we are left with 129 objects. We use
the version 1.2 of his code. The figure shows N/O vs O/H as obtained by
the \citet{PerezMontero.2014a} code on the left, and the comparison of
the O/H and N/O values to ones obtained by \bond on the middle and right
panels. The two rows correspond to two different runs of the
\citet{PerezMontero.2014a} code. On the top row, we have withheld the
\Oiiit line from the code, to see how it would behave using only strong
lines. This is not the recommended way of running the
\citet{PerezMontero.2014a} code, but this exercise shows that it cannot
be used as a strong line method.  The bottom row shows the results from
the \citet{PerezMontero.2014a} code when asking it to fit the \Oiiit
line as well (with a very strong weight as resulting from his
eq.~19). Note that the \citet{PerezMontero.2014a} code finds
systematically lower values of O/H than the \bond method, exactly like
the temperature-based method. The N/O is also rather scattered.

\begin{figure}
\centering
\includegraphics[width=\columnwidth, trim=10 10 10 10]{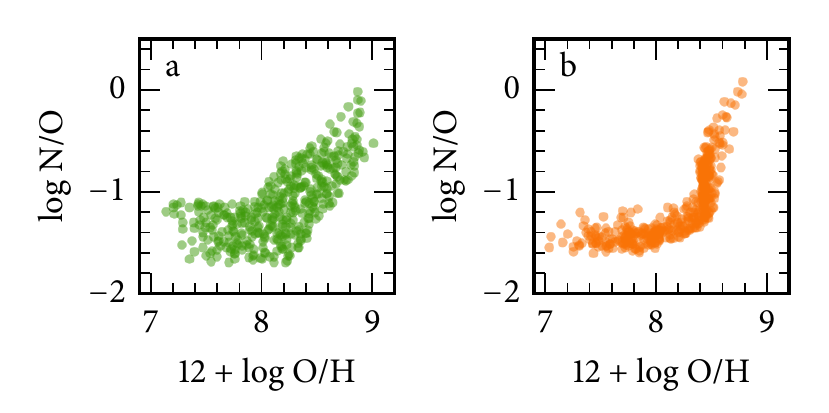}
\caption{(a) The original N/O vs O/H diagram for our fake sources, and
    (b) the N/O vs O/H diagram for our fake sources as calculated by
  the ON calibration from \citet{Pilyugin.Vilchez.Thuan.2010a}.}
\label{fig:fakeP10}
\label{lastpage} 
\end{figure}

Finally, we use the fake sample of Fig.~\ref{fig:fake} to show how the
ON strong-line method of \citet{Pilyugin.Vilchez.Thuan.2010a} biases the
abundance results in the N/O vs O/H diagram. As seen in
Fig.~\ref{fig:fakeP10}, the ON calibration from
\citet{Pilyugin.Vilchez.Thuan.2010a} has considerably squeezed the broad
input relation.

\bsp	


\end{document}